# Global Structure-Constrained 4D prestack inversion

Qingwei Zhao[1,2], Xuehua Chen[1,2], Shengyi Wang[1,2], Cong Wang[1,2], Gang Zhao[1,2]

[1]*Chengdu University of Technology, State Key Laboratory of Oil & Gas Reservoir Geology and Exploitation, Chengdu 610059, China.* [2]*Chengdu University of Technology, Key Laboratory of Earth Exploration & Information Techniques of Ministry of Education, Chengdu 610059, China.*

E-mail: zhaoqwei001@163.com; chen_xuehua@163.com; wangsy_min@163.com; wangcong@stu.cdut.edu.cn; zg20201109@163.com.

ABSTRACT

Prestack seismic inversion often suffers from inherent ill-posedness and is significantly affected by complex factors such as the wavelets, parameter coupling, heterogeneous media, fault displacements, and stratigraphic variations, making it difficult for the conventional inversion methods to accurately recover subsurface information. To address these challenges, this paper proposes a global inversion method based on multi-dimensional matrix multiplication to optimize the inversion results under structure constraints. This method first represents the inversion model as a four-dimensional tensor representing spatial and parameter dimensions. By performing multi-dimensional matrix multiplication operations in various directions of the model, a multi-dimensional tensor of observed data is effectively constructed. Secondly, multi-dimensional regularization are introduced. By analyzing the impact of different dimensional constraints on the inversion results, targeted parameter selection and adjustment are achieved, significantly enhancing the stability of the inversion process and effectively suppressing noise. Finally, stratigraphic and fault constraints extracted from seismic interpretation are incorporated to improve the geological coherence and reasonableness of the inversion results. Both synthetic and field data experiments demonstrate that the proposed method effectively enhances the robustness of inversion in complex structural areas while maintaining high computational efficiency, validating its effectiveness in practical applications.

## INTRODUCTION

Seismic inversion plays a crucial role in quantitatively reconstructing subsurface structures and lithological properties from seismic wavefields. However, in geologically complex areas characterized by heterogeneous media, faults, and stratigraphic variations, conventional inversion methods often struggle to accurately delineate subsurface features due to their inherent ill-posedness and high nonlinearity. Therefore, effective inversion requires a balance between geological constraints and advanced regularization techniques to ensure a unified representation of structural continuity and stratigraphic variability, thereby enhancing the robust inference of subsurface properties (Aki and Richards, 1980; Hansen, 1998).

Conventional seismic inversion methods, such as least-squares inversion, frequently yield unstable and low-resolution solutions when adequate constraints are lacking. Regularization techniques address this by introducing prior information through penalty terms in the objective function, thereby constraining the solution space and improving the stability and accuracy of the inversion. For instance, Tikhonov regularization, by limiting the L2 norm of model parameters, aims to smooth the solution but can lead to the blurring of sharp interfaces. To mitigate this, sparse regularization methods, including L1-norm regularization or basis pursuit inversion, promote sparsity in model parameters, enabling better recovery of sharp geological boundaries and consequently yielding high-resolution inversion results (Zhang et al., 2013). Bayesian inversion theory offers a more comprehensive framework for seismic inversion, treating both model parameters and noise as random variables. It employs Bayes' theorem to integrate prior information and observed data, yielding the posterior probability distribution of model parameters (Buland and More, 2003). Within the Bayesian framework, selecting appropriate prior distributions is critical for inversion quality. Incorporating non-Gaussian prior distributions with heavy-tailed characteristics, such as the Cauchy distribution, can more effectively describe the sparsity and discontinuity of model parameters, thus improving the lateral resolution of inversion results and the ability to identify thin layers (Alemie and Sacchi, 2011). Furthermore, combining advanced optimization algorithms, such as the Fast Iterative Shrinkage-Thresholding Algorithm (FISTA), can further enhance the computational efficiency and accuracy of high-resolution seismic inversion (Pérez et al., 2013).

Pre-stack inversion leverages amplitude variations with offset (AVO) information from seismic gathers, allowing for the simultaneous inversion of multiple elastic parameters, including P-wave velocity, S-wave velocity, and density. This provides richer rock physics attributes for reservoir geophysical interpretation, leading to more accurate fluid identification and reservoir characterization (Russell et al., 2011). Due to its relatively smaller contribution to the Zoeppritz equations and higher susceptibility to data noise, the density parameter is often one of the most unstable in pre-stack AVO inversion. Integrating practical geological understanding and well log data to establish empirical relationships between elastic parameters, such as empirical formulas between P-wave velocity and density, can effectively improve the robustness of inverted parameters (Gardner et al., 1974; Chen et al., 2018). Utilizing more accurate Zoeppritz equations or their improved linearized approximations can reduce forward modeling errors, thereby enhancing inversion accuracy (Gholami, 2015; Xu et al., 2024; Xu et al., 2025). Additionally, rock-physics-based fluid identification theories can provide more direct fluid identification attributes and porosity reservoir descriptions for inversion (Yin et al., 2014; Zong et al., 2014).

Multichannel inversion methods fully exploit the lateral continuity of seismic data in space, improving inversion stability and noise immunity by simultaneously processing multiple seismic traces. The core idea is to assume certain geological or geophysical correlations between adjacent seismic traces, thereby introducing lateral constraints during the inversion process. For example, Multitrace Total Variation (TV) regularization effectively suppresses random noise while preserving the sharp features and lateral continuity of geological interfaces (Gholami, 2015; Hamid and Pidlisecky, 2015). Furthermore, combining structure-oriented minimum entropy multichannel inversion (Geng et al., 2024) can reduce noise while highlighting the boundary information of geological features. Robust estimation of structural orientation parameters and 2D/3D local anisotropic Tikhonov regularization also provide more refined constraints for multichannel inversion (Gholami, 2024). Multiparameter regularization in acoustic impedance inversion also offers a method for balancing the weights of different parameters (Li and Peng, 2017).

Structure-oriented regularization methods further integrate subsurface geological structural information into the inversion process to ensure high consistency between inversion results and true geological structures. These methods guide the design of regularization terms by utilizing structural information from seismic data (e.g., formation dip, fault locations), thereby preserving and enhancing the continuity and clarity of geological interfaces during inversion. For instance, Structure-Oriented Full Waveform Inversion (FWI) significantly improves the accuracy and resolution of velocity model updates by incorporating structural priors into full waveform inversion (Gebre et al., 2025; Wo et al., 2025; Yao et al., 2025). Total Variation Waveform

Inversion also plays a role in preserving structural features (Yong et al., 2018). In post-stack multichannel structure-oriented regularization methods, by analyzing the structural characteristics of seismic data, noise is effectively suppressed while maintaining lateral continuity, thereby improving the quality of inversion results (Zhang et al., 2022). Pre-stack structurally constrained inversion enhances the stability and reliability of inversion by imposing structure constraints on elastic parameters (Hamid et al., 2018; Xu et al., 2025; Zhang et al., 2022). Additionally, techniques such as structure-oriented initial model building (Huang et al., 2021), fast 3D spatial structure constraints (Yang et al., 2022), and nonlinear pre-stack seismic inversion using edge-preserving smoothing filters (Zhe and Hanming, 2013) all aim to optimize inversion results through structural information. Deep learning methods with structure constraints, such as Structure Constraint CNN (Song et al., 2023), and regularization methods guided by stratigraphy, sequences, and faults (Wu, 2017), represent the latest developments in structured inversion. Methods utilizing directional structure tensors have also been widely used to guide structural smoothing and improve the lateral continuity of inversion results (Wu and Janson, 2017). These approaches estimate local dip and azimuth directly from seismic amplitudes and construct structure-oriented regularization operators. The application of anisotropic Tikhonov regularization in full waveform inversion of sparse data (Gholami, 2025), and multidip structure-oriented filtering (Guo et al., 2025; Zhang et al., 2025), further enhance inversion performance in complex structural areas. Unlike conventional structure-oriented inversion that relies on structure tensors estimated from seismic amplitudes, the proposed approach directly incorporates interpreted horizons and faults as structural priors. This interpretation-driven constraint helps preserve structural discontinuities across faults while maintaining stratigraphic continuity along horizons, thereby improving the geological consistency of the inverted elastic parameters.

Conventional seismic inversion methods face major challenges in complex geological settings and noisy environments, including non-uniqueness, poor stability, noise sensitivity, and limited ability to incorporate geological prior information, which restricts their effectiveness in fine reservoir characterization. In complex 3D media, it remains critical to exploit structural information inherent in seismic data efficiently and accurately, balance data misfit with model regularity, and control the substantial computational cost.

To address these issues, we propose an innovative global structure-constrained prestack inversion (GSCI) framework that improves stability, resolution, and geological plausibility while maintaining computational efficiency. The framework represents the inversion model as a four-dimensional tensor that couples spatial coordinates with model parameters, enabling simultaneous inversion along multiple structural directions within a unified mathematical formulation. Within this global formulation, multidimensional regularization is introduced through multidimensional matrix operations to impose refined constraints across different dimensions, facilitating constraint selection (e.g., sparsity, smoothness, structural fidelity), easing regularization tuning, suppressing noise, and mitigating ill-posedness. Furthermore, key structural geometries, such as stratigraphic horizons and faults interpreted from seismic data, are integrated as guiding priors to strengthen structural coherence and geological realism, particularly in areas with complex fault interactions, where the method demonstrates robust performance with high computational efficiency.

## THEORY AND METHOD

### Global seismic forward modeling

The Zoeppritz equations describe the relationship between incidence angle, reflection angle, and P-wave and S-wave velocities on both sides of an interface. Due to their highly nonlinear nature, researchers have developed various linear approximations based on these equations. Among them, the Aki–Richards approximation (Aki and Richards, 1980) describes the linear relationship between PP reflectivity and P-wave, S-wave velocities, and density, and can be expressed as:

$$R_{pp}(\theta) = \frac{1}{2}\sec^2\theta\frac{\Delta V_P}{V_P} - \left(4\gamma^2\sin^2\theta\right)\frac{\Delta V_S}{V_S} + \frac{1}{2}\left(1-4\gamma^2\sin^2\theta\right)\frac{\Delta\rho}{\rho} \tag{1}$$

Where $\theta$ is the incidence angle, $V_P = (V_{P1}+V_{P2})/2$ , $V_S = (V_{S1}+V_{S2})/2$, $\rho = (\rho_1+\rho_2)/2$, $\Delta V_P = V_{P2} - V_{P1}$, $\Delta V_S = V_{S2} - V_{S1}$ and $\Delta\rho = \rho_2 - \rho_1$. $V_{P1}$, $V_{S1}$, and $\rho_1$ are, respectively, the P-wave velocity, S-wave velocity, and density of the upper layer. $V_{P2}$, $V_{S2}$, and $\rho_2$ are, respectively, the P-wave velocity, S-wave velocity, and density of the under layer. where $\gamma$ is the S- and P-velocity ratio ($\gamma = V_S / V_P$). Assuming that $\gamma = 0.5$ to simplify.

When the changes in elastic properties between the upper and lower layers are small, the unknown terms in equation 1 can be expressed as:

$$\begin{aligned} \frac{\Delta V_P}{V_P} &\approx \Delta\ln(V_P) \\ \frac{\Delta V_S}{V_S} &\approx \Delta\ln(V_S) \\ \frac{\Delta\rho}{\rho} &\approx \Delta\ln(\rho) \end{aligned} \tag{2}$$

Substituting equation 2 into equation 1, we get:

$$R_{pp}(\theta) = \frac{1}{2}\sec^2\theta\Delta\ln(V_P) - \left(4\gamma^2\sin^2\theta\right)\Delta\ln(V_S) + \frac{1}{2}\left(1-4\gamma^2\sin^2\theta\right)\Delta\ln(\rho) \tag{3}$$

Seismic records can be modeled as the convolution of a seismic wavelet with reflection coefficients:

$$d(\theta) = w * R_{pp}(\theta) \tag{4}$$

From equation 3 and equation 4, the forward modeling process can be written in the following matrix form:

$$\begin{bmatrix} d(\theta_1) \\ d(\theta_2) \\ \vdots \\ d(\theta_N) \end{bmatrix} = \begin{bmatrix} \mathbf{W} & & & \\ & \mathbf{W} & & \\ & & \ddots & \\ & & & \mathbf{W} \end{bmatrix} \begin{bmatrix} a(\theta_1) & b(\theta_1) & c(\theta_1) \\ a(\theta_2) & b(\theta_2) & c(\theta_2) \\ \vdots & \vdots & \vdots \\ a(\theta_N) & b(\theta_N) & c(\theta_N) \end{bmatrix} \begin{bmatrix} \mathbf{D} & & \\ & \mathbf{D} & \\ & & \mathbf{D} \end{bmatrix} \begin{bmatrix} m_p \\ m_s \\ m_\rho \end{bmatrix} \tag{5}$$

Where $a(\theta) = \frac{1}{2}\sec^2\theta$ , $b(\theta) = 4\gamma^2 \sin^2\theta$ , $c(\theta) = \frac{1}{2}\left(1 - 4\gamma^2 \sin^2\theta\right)$ , $m_p = \ln(V_P)$ , $m_s = \ln(V_S)$, $m_\rho = \ln(\rho)$. $\mathbf{W}$ is the wavelet convolution matrix, $\mathbf{D}$ is the difference matrix.

equation 5 written in multi-dimensional matrix computation form:

$$\begin{bmatrix} d(\theta_1) & d(\theta_2) & \cdots & d(\theta_N) \end{bmatrix} = \mathbf{WD}\begin{bmatrix} m_p & m_s & m_\rho \end{bmatrix} \begin{bmatrix} a(\theta_1) & a(\theta_2) & \cdots & a(\theta_N) \\ b(\theta_1) & b(\theta_2) & \cdots & b(\theta_N) \\ c(\theta_1) & c(\theta_2) & \cdots & c(\theta_N) \end{bmatrix} \tag{6}$$

Then the global synthesized record forward equation:

$$\mathbf{d} = \mathbf{WDmL} = \mathbf{WD} \times \mathbf{m} \times_4 \mathbf{L} \tag{7}$$

Where the calculation rules are detailed in Appendix A. The dimensions of each matrix in equation 7 are:

$$\begin{aligned} &\mathbf{m} \in \mathbb{R}^{N_z \times N_x \times N_y \times N_c} \\ &\mathbf{W} \in \mathbb{R}^{N_z \times N_z} \\ &\mathbf{D} \in \mathbb{R}^{N_z \times N_z} \\ &\mathbf{L} \in \mathbb{R}^{N_c \times N_n} \\ &\mathbf{d} \in \mathbb{R}^{N_z \times N_x \times N_y \times N_n} \end{aligned} \tag{8}$$

Here $N_z$, $N_x$, $N_y$, $N_c$, *and* $N_n$ denote the numbers of samples in the *z*, *x*, *y*, parameter, and angle directions, respectively. The model **m** is a 4D tensor, composed of *z*, *x*, *y* spatial dimensions and c parameter dimension, the seismic record **d** is a 4D tensor, composed of *z*, *x*, *y* spatial dimensions and angle dimension. **m** is composed of $\ln(V_P)$ 、 $\ln(V_S)$ 、 $\ln(\rho)$, so $N_c = 3$; generally, 3 stacked angle datasets are used, so $N_n = 3$.

At this point, with no regularization, the inversion equation:

$$\mathbf{m} = \arg\min \| \mathbf{d} - \mathbf{GmL} \|_2^2 \tag{9}$$

Where, $\mathbf{G} = \mathbf{WD}$.

**Vertical resolution constraint**

We consider constraints in different dimensions separately. For example, in post-stack inversion, there are no parameter domain correlations, and typically only the vertical direction is considered, primarily the effect of the seismic wavelet. Therefore, effective vertical regularization constraints can effectively improve the resolution of the inversion.

Conventional low-frequency model constraint:

$$\hat{\mathbf{m}} = \arg\min \| \mathbf{d} - \mathbf{GmL} \|_2^2 + \lambda_1 \| \mathbf{m} - \mathbf{m_0} \|_2^2 \tag{10}$$

Writing the tensor **m** as a vector, and moving **G, L** to the left, equation 10 can be rewritten as

$$\hat{\mathbf{m}} = \arg\min \| \mathbf{d} - \tilde{\mathbf{G}}\mathbf{m} \|_2^2 + \lambda_1 \| \mathbf{m} - \mathbf{m_0} \|_2^2 \tag{11}$$

Where, $\tilde{\mathbf{G}} = \mathbf{L'} \otimes \mathbf{G}$ , $\otimes$ is the Kronecker Product symbol. This formulation corresponds to the commonly used objective function in conventional inversion methods.

Let $\mathbf{E} = \mathbf{mL}$, $\mathbf{E_0} = \mathbf{m}_0\mathbf{L}$, The objective function can then be expressed as:

$$\hat{\mathbf{E}} = \arg\min_{\mathbf{E}} \| \mathbf{d} - \mathbf{GE} \|_2^2 + \lambda_1 \| \mathbf{E} - \mathbf{E_0} \|_2^2 \tag{12}$$

In fact, the low-frequency model satisfies $\mathbf{E_0} = \mathbf{HE}$ , where **H** is the low-pass filter matrix. Therefore, the low-frequency constraint can be further improved as

$$\hat{\mathbf{E}} = \arg\min_{\mathbf{E}} \| \mathbf{d} - \mathbf{GE} \|_2^2 + \lambda_1 \| \mathbf{HE} - \mathbf{E_0} \|_2^2 \tag{13}$$

To further incorporate the reflection coefficient regularization constraint, the objective function becomes

$$\hat{\mathbf{E}} = \arg\min_{\mathbf{E}} \| \mathbf{d} - \mathbf{GE} \|_2^2 + \lambda_1 \| \mathbf{HE} - \mathbf{E_0} \|_2^2 + \lambda_2 \| \mathbf{DE} - \mathbf{R_0} \|_p^p \tag{14}$$

where $p$=1 or $p$=2. When $p$=1 for model data, and $p$=2 for real data, satisfying the sparse assumption of reflection coefficients or the Gaussian distribution assumption, respectively. $\mathbf{R_0}$ is the prior small reflection coefficient, generally 0.

The model parameter **m** can be obtained as follows:

$$\hat{\mathbf{m}} = \arg\min_{\mathbf{m}} \| \hat{\mathbf{E}} - \mathbf{mL} \|_2^2 = \hat{\mathbf{E}}\mathbf{L'}\left(\mathbf{LL'}\right)^{-1} \tag{15}$$

**Parameter-domain constraint**

According to equation 15, although the matrix is full rank, its condition number remains large, and its noise immunity is limited. Therefore, constraints need to be added to the parameter dimension. Gardner et al. (1974) proposed an empirical relationship between density and velocity, which can be written as

$$\rho \approx k V_P^a \tag{16}$$

The empirical relationship is added to the inversion constraint to improve the robustness of the inversion, and it is transformed into a logarithmic form as

$$a \ln(V_P) - \ln(\rho) + \ln(k) \approx 0 \tag{17}$$

By combining equation 17, equation 15 the inversion objective function with the empirical relationship constraint can be written as

$$\hat{\mathbf{m}} = \arg\min_{\mathbf{m}} \| \hat{\mathbf{E}} - \mathbf{mL} \|_2^2 + \beta \| (\mathbf{mT} - \mathbf{K}) \|_2^2 \tag{18}$$

Where, $\mathbf{T} = \begin{bmatrix} a \\ 0 \\ -1 \end{bmatrix}$, $\mathbf{K} = \begin{bmatrix} \ln(k) \\ \vdots \\ \ln(k) \end{bmatrix}$.

Another approach to improve the inversion stability of matrix $(\mathbf{LL'})$ is given by

$$\hat{\mathbf{m}} = \arg\min_{\mathbf{m}} \| \hat{\mathbf{E}} - \mathbf{mL} \|_2^2 + \beta \| \mathbf{m} \|_2^2 = \hat{\mathbf{E}}\mathbf{L'}(\mathbf{LL'} + \beta \mathbf{I})^{-1} \tag{19}$$

Since the inversion has different sensitivities to different parameters (P-wave velocity, S-wave velocity, and density), introducing different regularization parameters for each parameter is a more reasonable choice. The objective function can therefore be written as

$$\hat{\mathbf{m}} = \arg\min_{\mathbf{m}} \| \hat{\mathbf{E}} - \mathbf{mL} \|_2^2 + \| \mathbf{m}\boldsymbol{\beta}_{\mathbf{c}} \|_2^2 = \hat{\mathbf{E}}\mathbf{L'}(\mathbf{LL'} + \boldsymbol{\beta}_{\mathbf{c}}\boldsymbol{\beta}_{\mathbf{c}}')^{-1} \tag{20}$$

Where $\boldsymbol{\beta}_{\mathbf{c}} = \begin{bmatrix} \beta_1 & 0 & 0 \\ 0 & \beta_2 & 0 \\ 0 & 0 & \beta_3 \end{bmatrix}$.

**Lateral constraint**

Multichannel inversion methods utilize information from adjacent traces to improve the accuracy of inversion. The inversion result for each seismic trace is a comprehensive result from adjacent traces. The Multitrace Total Variation regularization method is commonly used (Gholami, 2015; Hamid and Pidlisecky, 2015). If only the lateral regularization constraint is considered, the objective function can be written as

$$\begin{aligned} J(m) &= \min_{\mathbf{m}} \| \mathbf{d} - \mathbf{GmL} \|_2^2 \\ s.t. \quad & \mathbf{mD_x} = 0, \\ & \mathbf{mD_y} = 0 \end{aligned} \tag{21}$$

The corresponding inversion result can be expressed as

$$\hat{\mathbf{m}} = \arg\min_{\mathbf{m}} \| \mathbf{d} - \mathbf{G} \times \mathbf{m} \times_4 \mathbf{L} \|_2^2 + \alpha_1 \| \mathbf{m} \times_2 \mathbf{D_x} \|_p^p + \alpha_2 \| \mathbf{m} \times_3 \mathbf{D_y} \|_p^p \tag{22}$$

Where $p$ is selected as $p$=1 or $p$=2 based on the lateral continuous sparsity. $\mathbf{D_x}$ is the first-order difference matrix in the x direction, and $\mathbf{D_y}$ is the first-order difference matrix in the y direction. When $p$=1, the alternating direction method of multipliers (ADMM) algorithm is used for solving; when $p$=2, the Conjugate Gradient iterative method is used for solving.

When $\alpha_1, \alpha_2 << 1$, $p = 2$, applying multi-dimensional matrix multiplication to the data tensor d yields an approximation of the true model. As detailed in Appendix B, the inversion result is expressed as:

$$\hat{\mathbf{m}} \approx \left[ \left[ \left(\mathbf{G'G}\right)^{-1} \mathbf{G'} \right] \mathbf{d} \times_4 \left[ \mathbf{L'} \left(\mathbf{LL'}\right)^{-1} \right] \right] \times_2 \left[ \left( \mathbf{I} + \alpha_1 \mathbf{D_x} \mathbf{D_x'} \right)^{-1} \right] \times_3 \left[ \left( \mathbf{I} + \alpha_2 \mathbf{D_y} \mathbf{D_y'} \right)^{-1} \right] \tag{23}$$

**Structure constraint**

Multichannel inversion with lateral constraints, in complex areas where horizons are not horizontal, often uses structural tensors to obtain main directions to improve lateral constraints and achieve better structural inversion results (Yang et al., 2022; Gholami, 2024). This study uses structural information such as strata and faults obtained from seismic interpretation to enhance the effect of lateral regularization constraints and improve inversion quality.

*Layer flattening transformation (phase interpolation technique)*

We assume there is a matrix that, when multiplied by the model, can flatten the model of curved horizons, and its inverse matrix restores the flattened model to its original state. The relationship can be expressed as

$$\begin{aligned}\mathbf{m}_{\mathbf{T}} &= \mathbf{T}\mathbf{m} \\ \mathbf{m} &= \mathbf{T}^{-1}\mathbf{m}_{\mathbf{T}}\end{aligned} \tag{24}$$

Because horizon information is usually limited, the interpreted horizons are used as control points. Assuming two horizons as control points, the flattening transformation can be written as

$$\mathbf{m}_{\mathbf{T}} = \mathbf{T}\mathbf{m} = \begin{bmatrix} ? & 1 & ? & 0 \\ ? & 0 & ? & 0 \\ ? & 0 & ? & 1 \\ ? & 0 & ? & 0 \end{bmatrix} \begin{bmatrix} \\ h_1 \\ \\ h_2 \end{bmatrix} \tag{25}$$

Where $h_1$, $h_2$ represent Layer 1, Layer 2, $?$ denotes the missing elements of matrix **T**.

To construct the complete matrix **T**, we observe that each column of **T** has the same amplitude in the frequency domain while the phase varies smoothly. Therefore, a Fourier transform is applied to matrix **T** to separate it into amplitude and phase components. The phase component is then reconstructed using interpolation techniques, after which an inverse Fourier transform is performed to obtain the completed matrix **T**:

$$\mathbf{F}\mathbf{T} = \mathbf{T}^{\mathbf{a}} e^{i\mathbf{T}^{\mathbf{\Phi}}} \tag{26}$$

Where, $\mathbf{F}$ denotes the Fourier transform matrix, $\mathbf{T}^{\mathbf{a}}$ represents the amplitude spectrum of **T**, and $\mathbf{T}^{\mathbf{\Phi}}$ represents the phase spectrum of **T**.

The matrix **T** can therefore be reconstructed using

$$\begin{aligned}&\hat{\mathbf{T}}^{\mathbf{\Phi}} = \arg\min_{\mathbf{T}^{\mathbf{\Phi}}} \| \mathbf{T}^{\mathbf{\Phi}}\mathbf{S} - \mathbf{T}^{\mathbf{\Phi}}_{\mathbf{S}} \|_2^2 + \lambda \| \mathbf{T}^{\mathbf{\Phi}}\mathbf{D}_{\mathbf{x}} \|_2^2 \\ &\mathbf{T} = \mathbf{F}^{-1}\left(\mathbf{T}^{\mathbf{a}} e^{i\hat{\mathbf{T}}^{\mathbf{\Phi}}}\right)\end{aligned} \tag{27}$$

The inverse matrix $\mathbf{T}^{-1}$ satisfies

$$\mathbf{m} = \mathbf{T}^{-1}\mathbf{m}_{\mathbf{T}} = \arg\min_{\mathbf{m}} \| \mathbf{m}_{\mathbf{T}} - \mathbf{T}\mathbf{m} \|_2^2 + \lambda \| \mathbf{D}\mathbf{m} \|_2^2 \quad (28)$$

$$\mathbf{T}^{-1} \approx \left(\mathbf{T}'\mathbf{T} + \lambda\mathbf{D}'\mathbf{D}\right)^{-1}\mathbf{T}' \quad (29)$$

By incorporating the layer-flattening technique, equation 22 is modified to:

$$\hat{\mathbf{m}} = \arg\min_{\mathbf{m}} \| \mathbf{d} - \mathbf{G} \times \mathbf{m} \times_4 \mathbf{L} \|_2^2 + \alpha_1 \| \mathbf{T}_{\mathbf{N}}\mathbf{m} \times_2 \mathbf{D}_{\mathbf{x}} \|_p^p + \alpha_2 \| \mathbf{T}_{\mathbf{N}}\mathbf{m} \times_3 \mathbf{D}_{\mathbf{y}} \|_p^p \quad (30)$$

*Fault-guided regularization*

In real subsurface environments, geological structures often include not only curved strata but also faults. At fault locations, the model becomes discontinuous along the stratigraphic direction. Some existing approaches enforce layer continuity separately within the hanging wall and footwall of faults; however, this increases the complexity of the constraints. To address this issue, a fault-guided regularization constraint is introduced, and equation 30 can be simplified as

$$\hat{\mathbf{m}} = \arg\min_{\mathbf{m}} \| \mathbf{d} - \mathbf{G}\mathbf{m}\mathbf{L} \|_2^2 + \alpha_1 \| [\mathbf{T}_{\mathbf{N}}\mathbf{m}\mathbf{D}_{\mathbf{x}}] \odot \mathbf{K}_{\mathbf{x}} \|_p^p + \alpha_2 \| [\mathbf{T}_{\mathbf{N}}\mathbf{m}\mathbf{D}_{\mathbf{y}}] \odot \mathbf{K}_y \|_p^p \quad (31)$$

Where $\mathbf{K}_{\mathbf{x}} \in \mathbb{R}^{N_z \times N_x \times N_y \times N_c}$ is the x-direction fault tensor, which is 0 at fault locations and 1 elsewhere; $\mathbf{K}_{\mathbf{y}} \in \mathbb{R}^{N_z \times N_x \times N_y \times N_c}$ is the y-direction fault tensor, which is 0 at fault locations and 1 elsewhere. The symbol $\odot$ is the Hadamard product, meaning element-wise multiplication. Using $p$=1 is to retain large gradients and discard small gradients in the first-order lateral gradient domain, while $\mathbf{K}_\mathbf{x}$ and $\mathbf{K}_\mathbf{y}$ are already explicitly given. Therefore, when there is fault-guided regularization, $p$=2 is sufficient.

The final global structure-constrained 4D pre-stack inversion objective function can therefore be written as

$$\hat{\mathbf{m}} = \arg\min_{\mathbf{m}} \| \mathbf{d} - \mathbf{G}\mathbf{m}\mathbf{L} \|_2^2 + \alpha_1 \| [\mathbf{T}_{\mathbf{N}}\mathbf{m}\mathbf{D}_{\mathbf{x}}] \odot \mathbf{K}_{\mathbf{x}} \|_2^2 + \alpha_2 \| [\mathbf{T}_{\mathbf{N}}\mathbf{m}\mathbf{D}_{\mathbf{y}}] \odot \mathbf{K}_y \|_2^2 \quad (32)$$

Using the inversion results obtained after applying the vertical and parameter constraints as the initial model, equation 32 is solved iteratively as

$$\begin{cases} \mathbf{m_T} = \mathbf{T_N m} \\ \mathbf{\hat{m}_T} = \arg\min_{\mathbf{m_T}} \| \mathbf{d} - \mathbf{G T_N^{-1} m_T L} \|_2^2 + \alpha_1 \| \mathbf{m_T D_x} \odot \mathbf{K_x} \|_2^2 + \alpha_2 \| \mathbf{m_T D_y} \odot \mathbf{K_y} \|_2^2 \\ \mathbf{\hat{m}} = \mathbf{T_N^{-1} \hat{m}_T} \end{cases} \tag{33}$$

with the iteration formula given by

$$\mathbf{\hat{m}_T^{k+1}} = \mathbf{\hat{m}_T^k} - \zeta \left[ 2\mathbf{T_N G'} \left( \mathbf{G T_N^{-1} \hat{m}_T^k L} - \mathbf{d} \right) \mathbf{L'} + 2\alpha_1 \left( \mathbf{\hat{m}_T^k D_x} \odot \mathbf{K_x} \right) \mathbf{D_x'} + 2\alpha_2 \left( \mathbf{\hat{m}_T^k D_y} \odot \mathbf{K_y} \right) \mathbf{D_y'} \right] \tag{34}$$

By combining equation 14, equation 20, and equation 33, the workflow of the global structure-constrained 4D pre-stack inversion can be summarized as follows.

**Algorithm 1. Global Structure-constrained Prestack Inversion (GSCI)**

---

**Vertical dimension:**

1) Set $\mathbf{G} = \mathbf{WD}$, $\mathbf{E} = \mathbf{mL}$, Estimate the practical low-frequency filter **H**, and determine regularization factors $\lambda_1$, $\lambda_2$.

2)According to the sparsity of the reflection coefficient, set *p*=2, and solve $\mathbf{\hat{E}} = \arg\min_{\mathbf{E}} \| \mathbf{d} - \mathbf{GE} \|_2^2 + \lambda_1 \| \mathbf{HE} - \mathbf{E_0} \|_2^2 + \lambda_2 \| \mathbf{DE} - \mathbf{R_0} \|_p^p$. a least-squares direct inversion is used.

**Parameter dimension:**

3) Based on the influence of parameters $V_P$, $V_S$ and $\rho$, determine regularization $\boldsymbol{\beta}_\mathbf{c}$.

4) $\mathbf{\hat{m}} = \arg\min_{\mathbf{m}} \| \mathbf{\hat{E}} - \mathbf{mL} \|_2^2 + \| \mathbf{m}\boldsymbol{\beta}_\mathbf{c} \| \|_2^2 = \mathbf{\hat{E}L'} \left( \mathbf{LL'} + \boldsymbol{\beta}_\mathbf{c} \boldsymbol{\beta}_\mathbf{c}' \right)^{-1}$

**Structure constraint (lateral dimension):**

5) Establish the global structural flattening transform $\mathbf{T_N}$ and its inverse $\mathbf{T_N^{-1}}$ .Construct fault-guided tensor $\mathbf{K_x}$, $\mathbf{K_y}$. and differential operators $\mathbf{D_x}$, $\mathbf{D_y}$ for x and y directions.

6) $\mathbf{\hat{m}_T} = \mathbf{T_N \hat{m}}$

7) Choose step size $\zeta$ and regularization parameters $\alpha_1$, $\alpha_2$

8) Use the gradient descent method to solve

$$\mathbf{\hat{m}_T} = \arg\min_{\mathbf{m_T}} \| \mathbf{d} - \mathbf{G T_N^{-1} m_T L} \|_2^2 + \alpha_1 \| \mathbf{m_T D_x} \odot \mathbf{K_x} \|_2^2 + \alpha_2 \| \mathbf{m_T D_y} \odot \mathbf{K_y} \|_2^2$$

Initialize: $\mathbf{\hat{m}_T^0} = \mathbf{\hat{m}_T}$.

While $| \hat{m}_T^{k+1} - \hat{m}_T^k \|_2^2 > \varepsilon$, perform the iterative update:

$$\mathbf{\hat{m}_T^{k+1}} = \mathbf{\hat{m}_T^k} - \zeta \left[ 2\mathbf{T_N G'} \left( \mathbf{G T_N^{-1} \hat{m}_T^k L} - \mathbf{d} \right) \mathbf{L'} + 2\alpha_1 \left( \mathbf{\hat{m}_T^k D_x} \odot \mathbf{K_x} \right) \mathbf{D_x'} + 2\alpha_2 \left( \mathbf{\hat{m}_T^k D_y} \odot \mathbf{K_y} \right) \mathbf{D_y'} \right]$$

End

9) $\mathbf{\hat{m}} = \mathbf{T_N^{-1} \hat{m}_T}$

10) $[V_P \quad V_S \quad \rho] = \exp(\hat{\mathbf{m}})$

## NUMERICAL EXAMPLES

We first evaluate the GSCI using the bottom-right region of the 2D Marmousi-2 model to test the partial regularization constraints. The model contains 1001 CDPs with 801 sampling points per trace and a sampling interval of 1 ms. Synthetic seismograms at incident angles of 10°, 20°, and 30° are generated using a 20 Hz Ricker wavelet, and Gaussian white noise with a noise-to-signal (N/S) ratio of 0.1 is added (Figure 1).

In the noise-free case, the effect of the vertical constraint (equation 14) is first examined. The inversion results for $p$=2 and $p$=1 are shown in Figure 2. The trace comparison at CDP 201 indicates that when $p$=1, the true P-wave velocity, S-wave velocity, and density are accurately recovered. In contrast, the $p$=2 result generally follows the true model but lacks high-frequency components and shows slight oscillations at layer interfaces (Figure 3). Although L1-norm regularization preserves large reflection coefficients while suppressing small ones, small coefficients may be incorrectly removed in practice. Therefore, the L2-norm is adopted in subsequent experiments.

Under noisy conditions, the effect of parameter-domain regularization is further investigated. Figure 4 compares the inversion results obtained using the conventional algorithm (equation 11) and the global inversion with specific regularization coefficients (equation 19). Because different parameters contribute differently to forward modeling, their robustness to noise varies: P-wave velocity is recovered most accurately, followed by S-wave velocity, while density shows the poorest accuracy.

The parameter-domain regularization is further improved by introducing the global inversion with optimized coefficients (equation 20) together with the P-wave velocity–density empirical relationship (equation 18). The resulting inversion (Figure 5) shows significant improvement, especially for density. The trace comparison at CDP 201 confirms that the conventional algorithm and equation (19) yield similar results, whereas the introduction of equation (20) and the empirical constraint markedly enhances the density inversion (Figure 6).

Figure 7 shows the layer-flattening matrix constructed from 14 interpreted horizons, the completed matrix obtained using phase interpolation (equation 27), its inverse matrix (equation 29), and the effectiveness of the flattening process. In the presence of noise, the layer-flattening technique is applied to improve the lateral regularization constraint. Figure 8 compares the conventional laterally constrained global inversion (equation 22) with the layer-flattened inversion (equation 30). The trace comparison at CDP 201 indicates that lateral constraints reduce noise, while the layer-flattened inversion further improves structural clarity, particularly near horizon boundaries (Figure 9).

To evaluate the overall performance of the GSCI, a synthetic 3D overthrust model containing P-wave velocity, S-wave velocity, and density is used. A time slice at 230 ms is selected for comparison (Figure 10). The model contains 201 Inlines and 201 Crosslines with 140 samples per trace and a sampling interval of 5 ms. Synthetic seismograms at incident angles of 10°, 20°, and 30° are generated using a 20 Hz Ricker wavelet with Gaussian white noise added at an N/S ratio of 0.3. The model includes both fault planes and relatively flat strata, providing a suitable test for structural constraints.

Four inversion strategies are compared: global inversion without TV regularization, global inversion with L2-TV regularization (equation 22), global inversion with L1-TV regularization (equation 22), and the proposed structurally constrained global inversion (equation 33) (Figures 11–12). Time slices at 230 ms (Figures 13–14) show that the inversion without TV regularization is strongly affected by noise. The L2-TV inversion reduces noise and produces smooth results but blurs structural boundaries. The L1-TV inversion improves edge preservation while suppressing noise. In contrast, the structurally constrained inversion achieves both strong noise suppression and the best boundary preservation.

Vertical profiles at Inline = 100 (Figures 15–16) further confirm these observations. The unregularized result contains strong noise; the L2-TV result reduces noise but blurs boundaries; the L1-TV result preserves boundaries better; and the structurally constrained inversion provides the clearest structures with minimal noise.

Finally, single-trace comparisons at Inline = 100 and Crossline = 50 show that the unconstrained and L2-TV inversions suffer from significant noise and blurred interfaces, while the L1-TV inversion improves edge definition. The structurally constrained global inversion produces results closest to the true model (Figure 17).

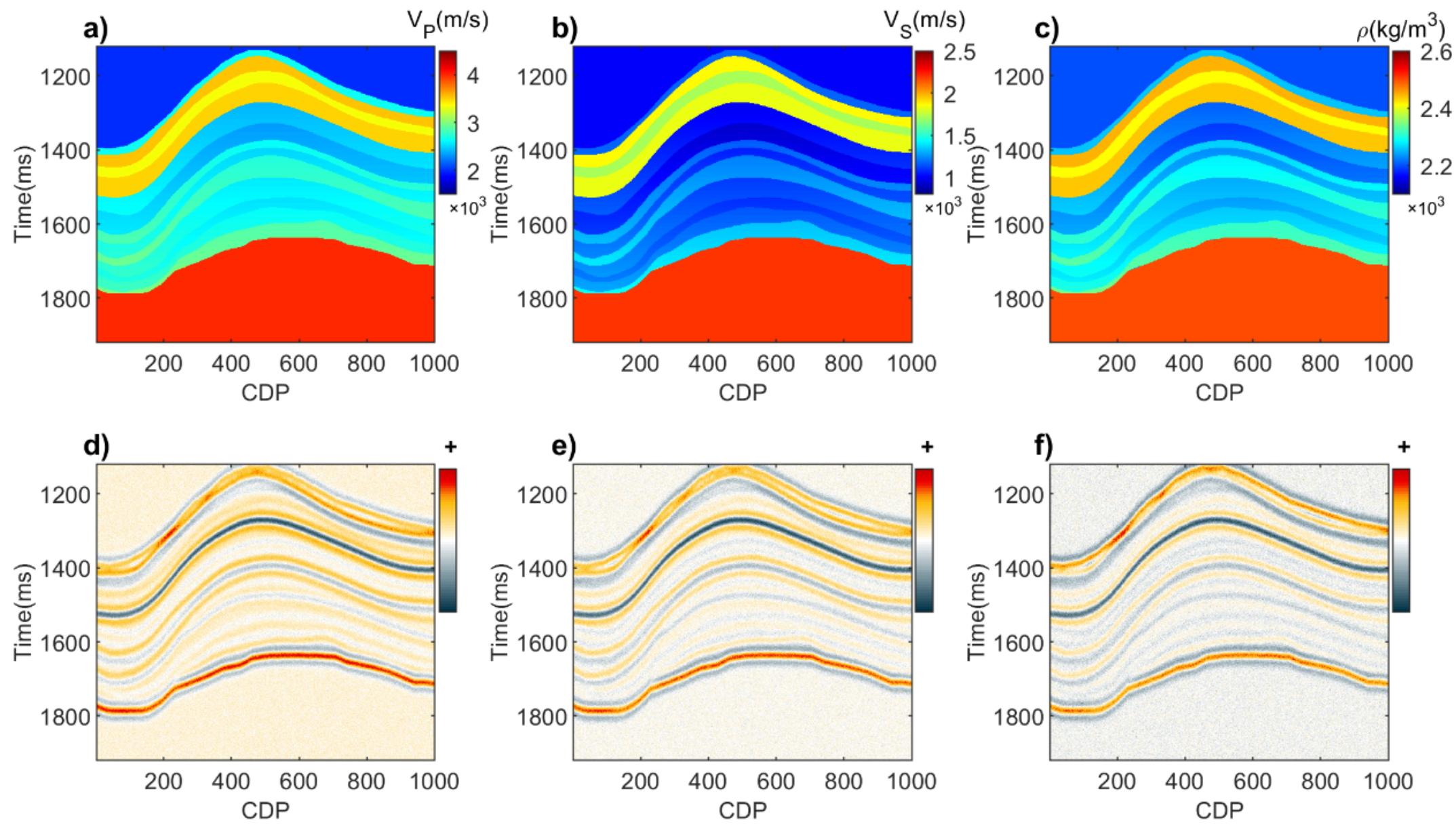

Figure 1. The lower-right part of the Marmousi-2 model of the (a) P-wave velocity, (b) S-wave velocity, (c) density. The noisy seismic gathers with angles (d) 10°, (e) 20°, and (f) 30°.

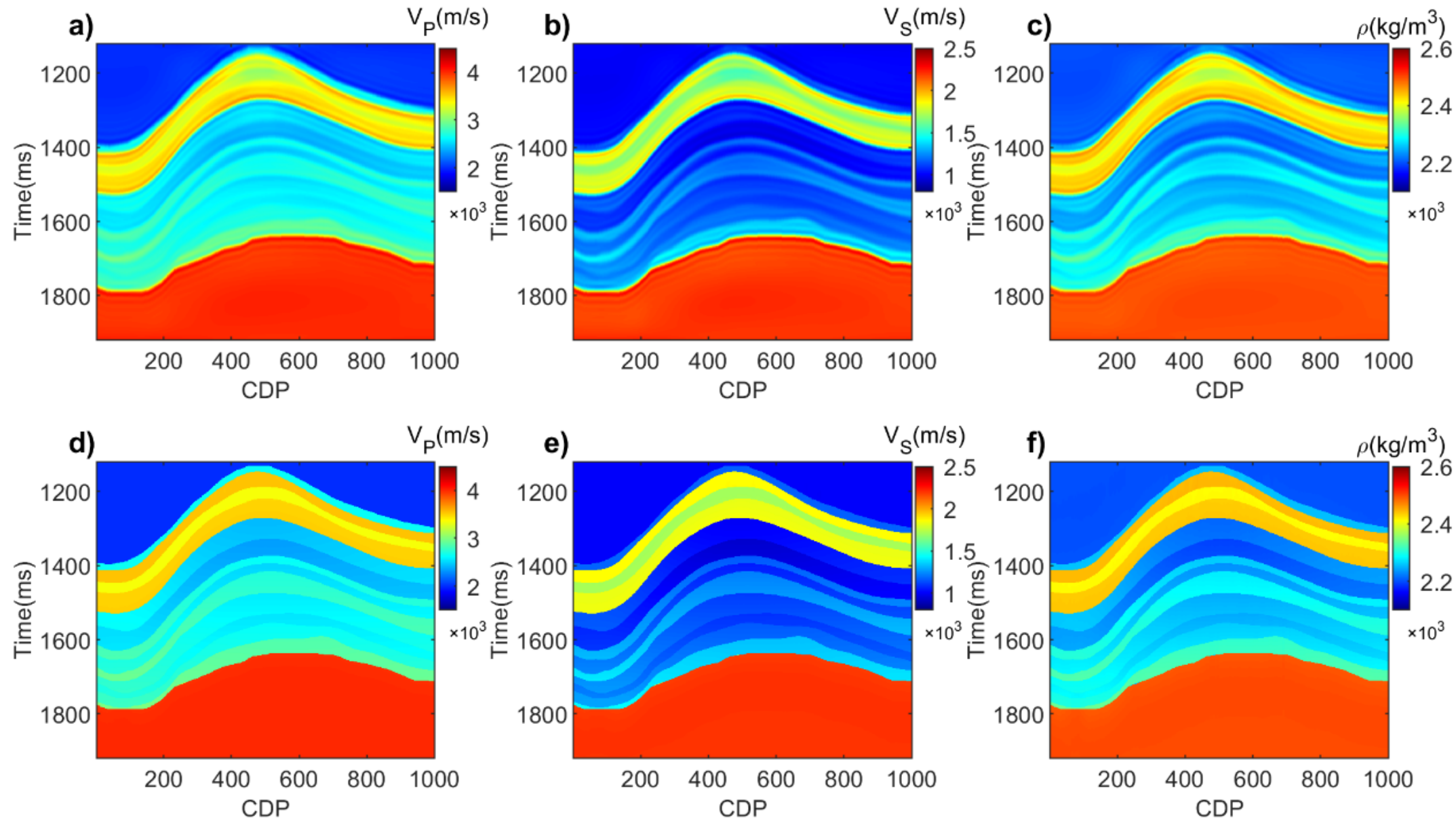

Figure 2. The estimated (a) P-wave velocity, (b) S-wave velocity, and (c) density with p=2 under noise-free conditions; The estimated (d) P-wave velocity, (e) S-wave velocity, and (f) density with p=1 under noise-free conditions.

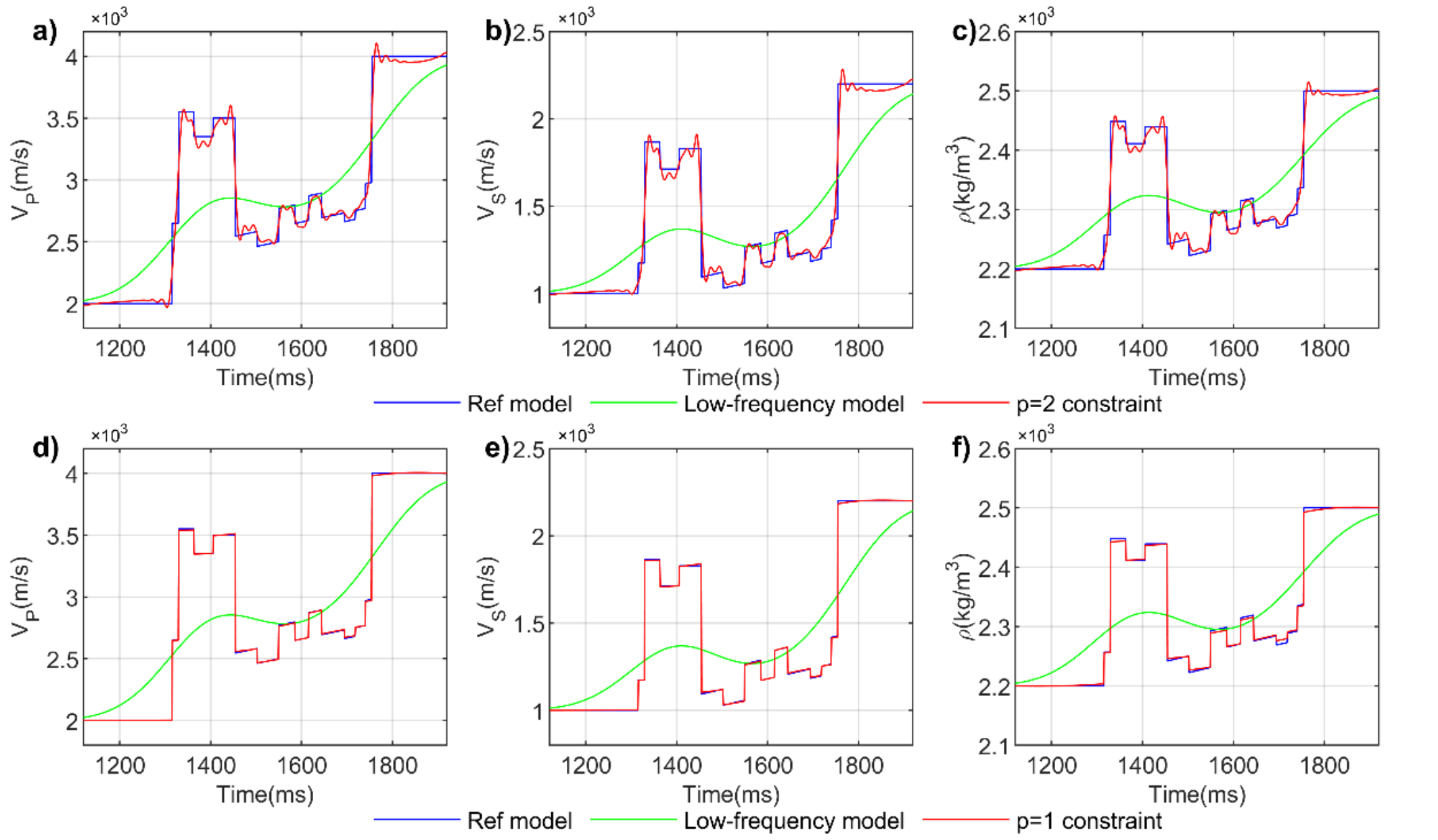

Figure 3. The estimated (a) P-wave velocity, (b) S-wave velocity, and (c) density with p=2 at CDP 201 under noise-free conditions; The estimated (d) P-wave velocity, (e) S-wave velocity, and (f) density with p=1 at CDP 201 under noise-free conditions.

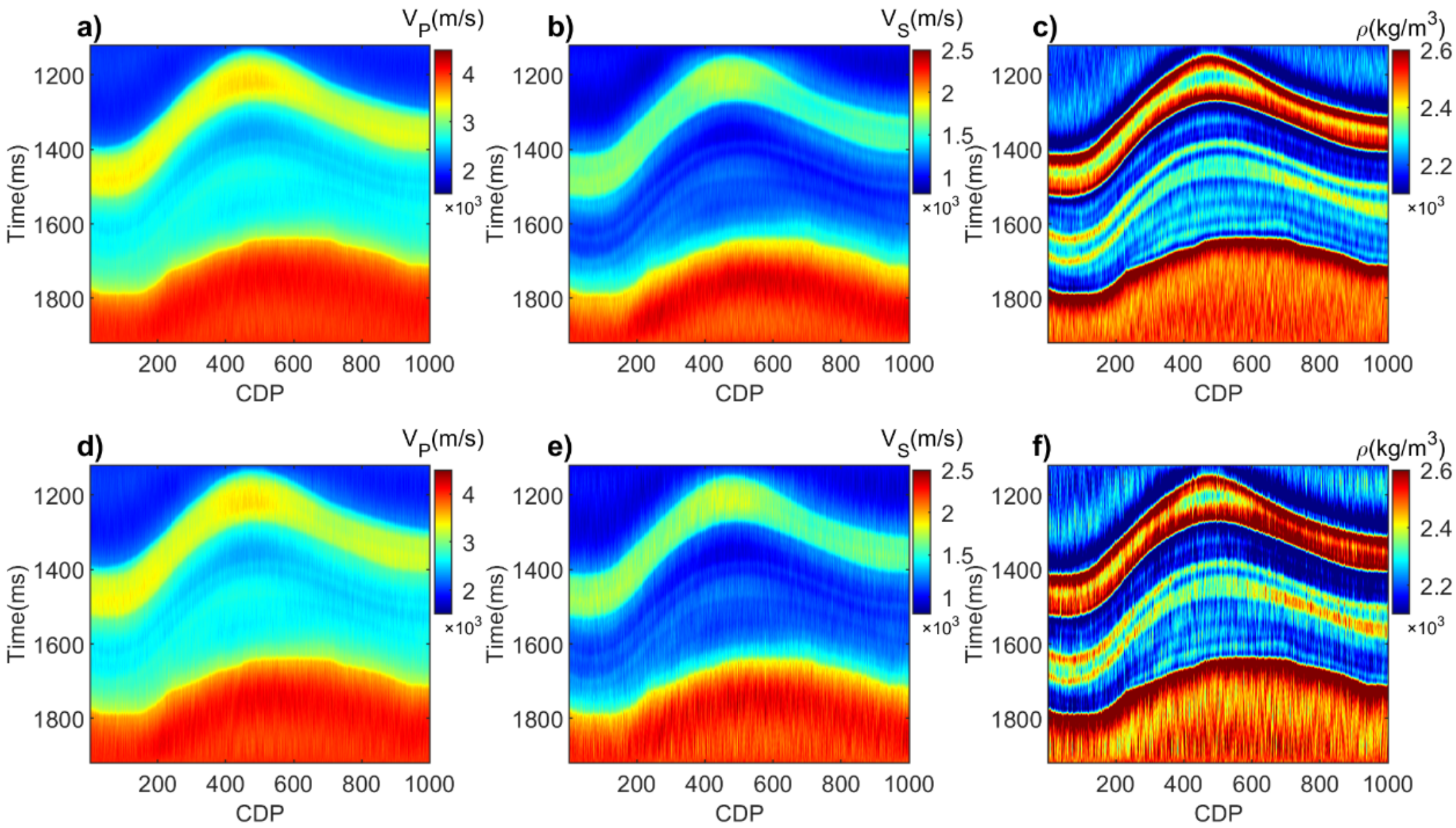

Figure 4. The estimated (a) P-wave velocity, (b) S-wave velocity, and (c) density by regular algorithm; The estimated (d) P-wave velocity, (e) S-wave velocity, and (f) density with global algorithm $\beta$.

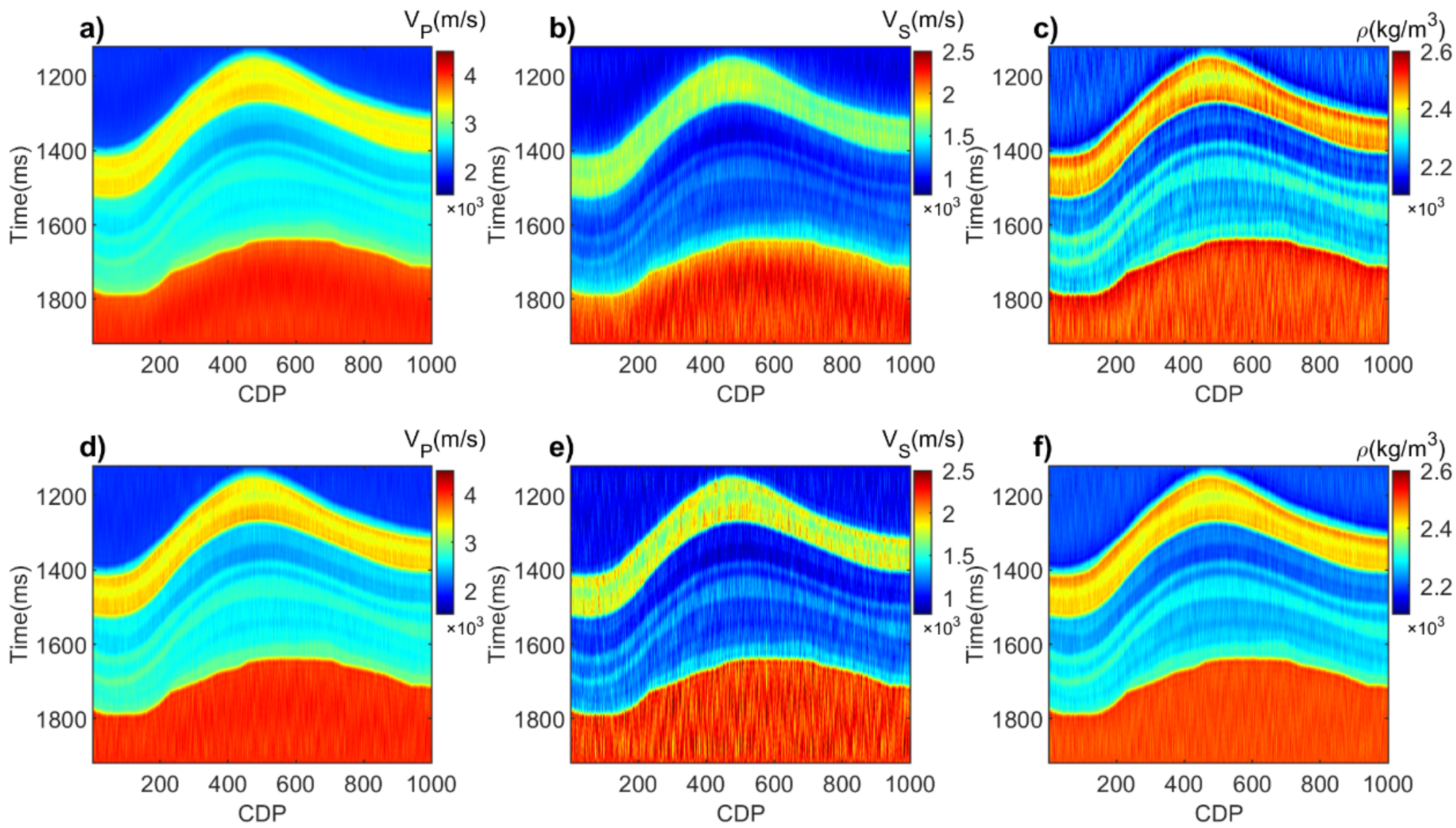

Figure 5. The estimated (a) P-wave velocity, (b) S-wave velocity, and (c) density by global algorithm $\boldsymbol{\beta}_c$; The estimated (d) P-wave velocity, (e) S-wave velocity, and (f) density by global experience constraints.

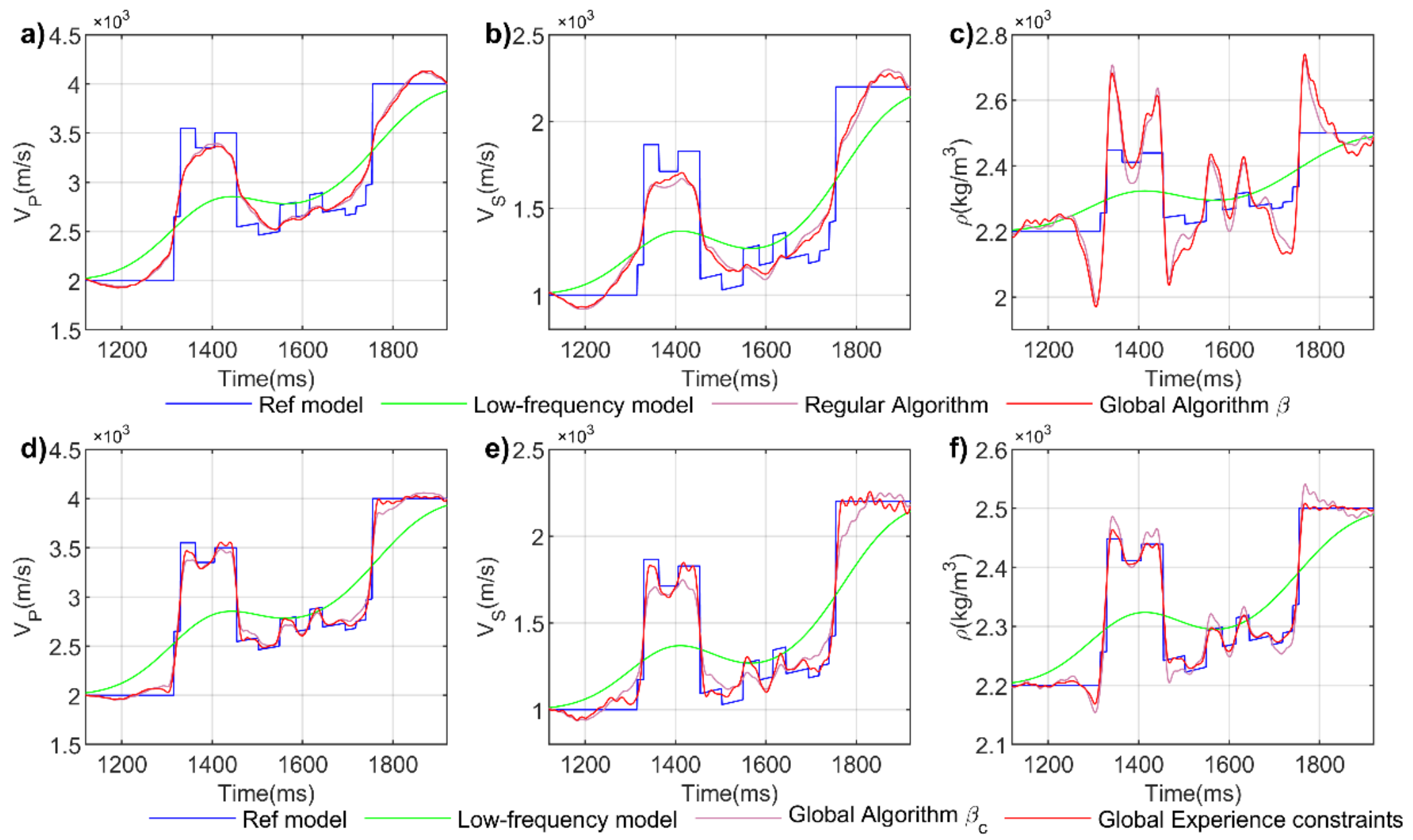

Figure 6. The estimated (a) P-wave velocity, (b) S-wave velocity, and (c) density by regular algorithm and global algorithm $\beta$ at CDP 201; The estimated (d) P-wave velocity, (e) S-wave velocity, and (f) density by global algorithm $\boldsymbol{\beta}_c$ and global experience constraints at CDP 201.

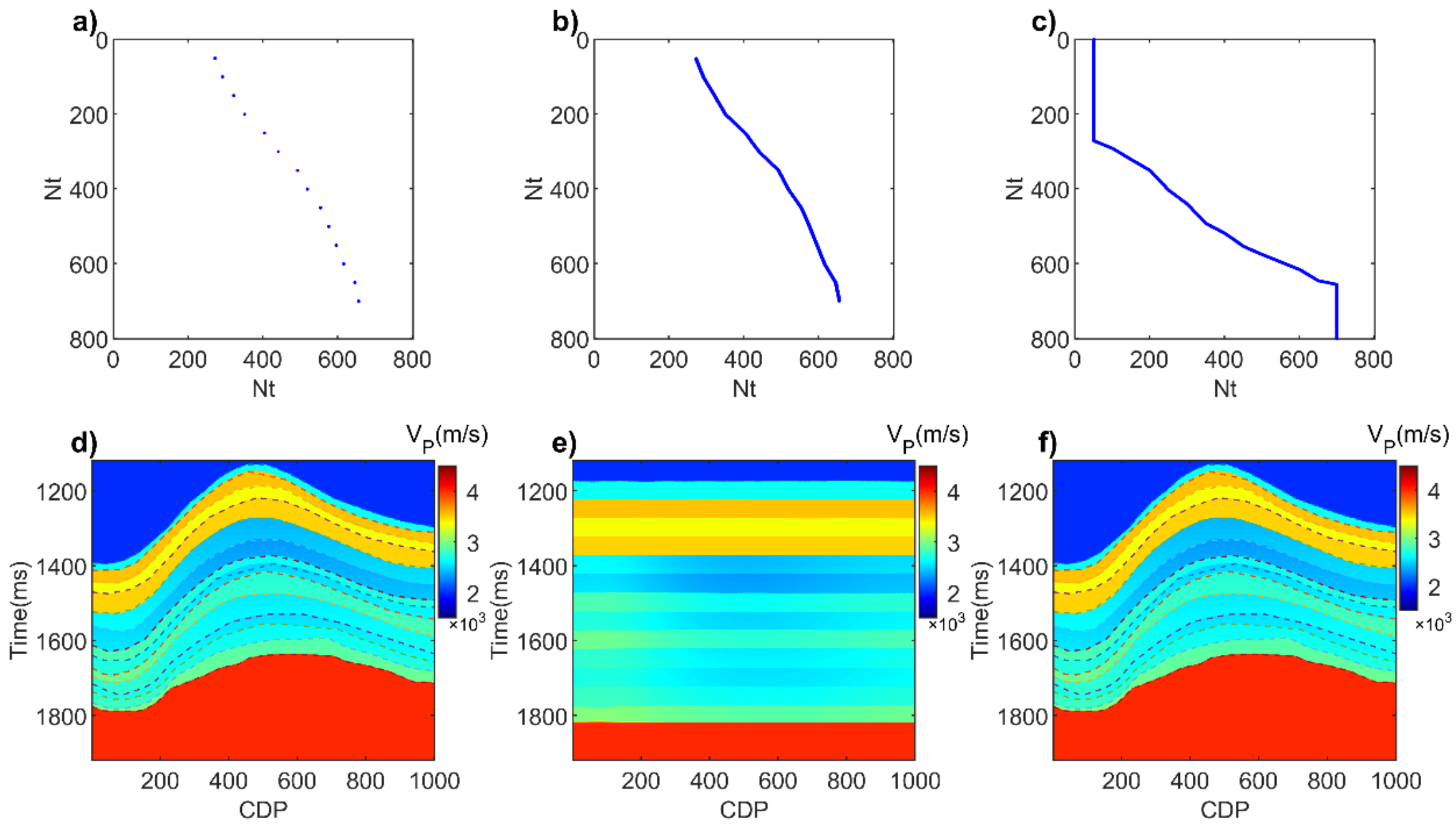

Figure 7. First trace (a) layer-flattening matrix (uninterpolated), (b) layer-flattening matrix (after interpolation), (c) inverse layer-flattening matrix; (d) original model, (e) model after layer-flattening, (f) model after inverse layer-flattening.

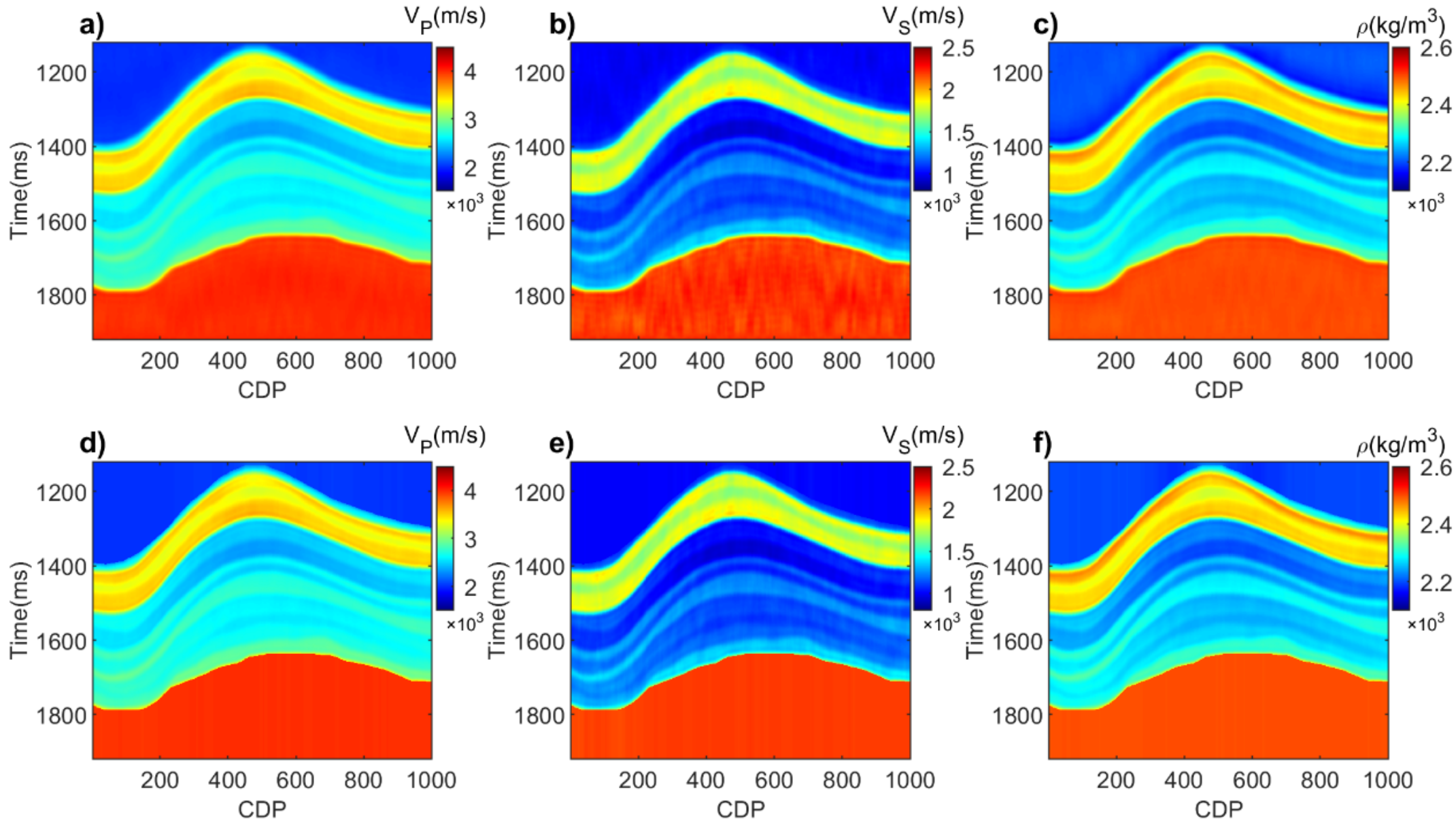

Figure 8. The estimated (a) P-wave velocity, (b) S-wave velocity, and (c) density by Lateral constraint global algorithm . The estimated (d) P-wave velocity, (e) S-wave velocity, and (f) density by layer- flattening lateral constraint global algorithm.

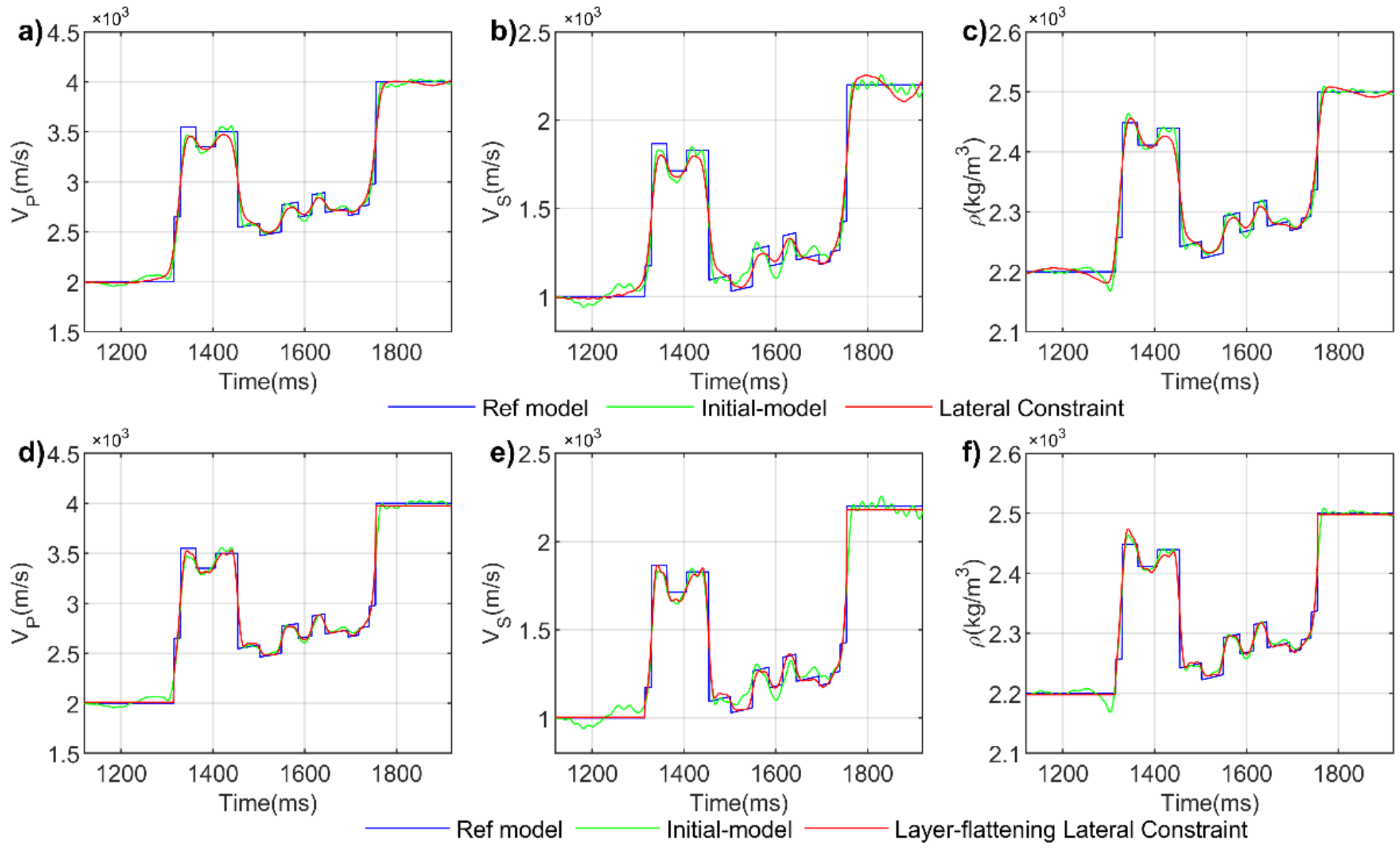

Figure 9. The estimated (a) P-wave velocity, (b) S-wave velocity, and (c) density by Lateral constraint global algorithm at CDP 201. The estimated (d) P-wave velocity, (e) S-wave velocity, and (f) density by layer- flattening lateral constraint global algorithm at CDP 201.

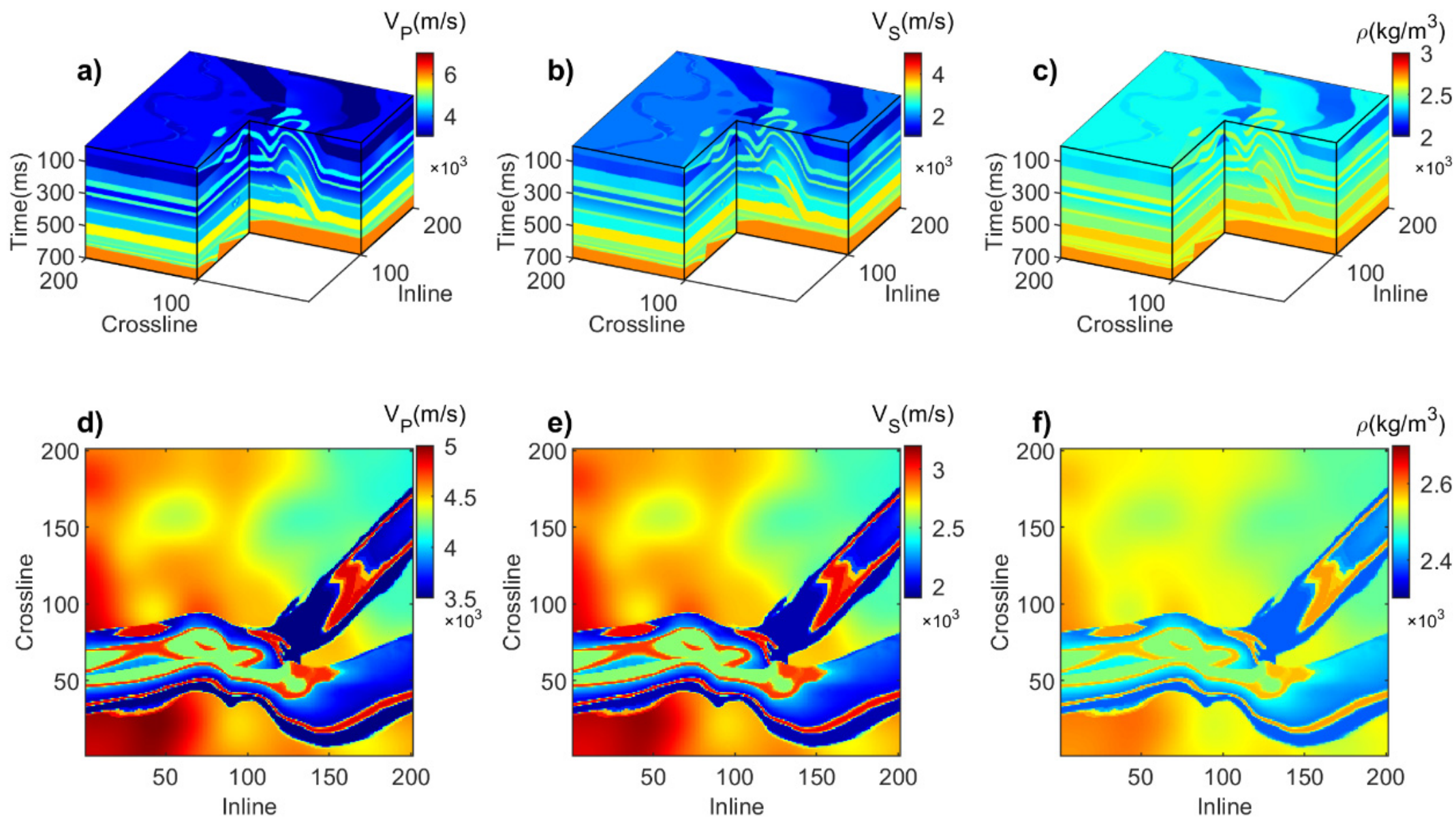

Figure 10: The 3D overthrust model of (a) P-wave velocity, (b) S-wave velocity, and (c) density. Figures (d), (e), and (f) show the 230 ms time slices of the P-wave velocity, S-wave velocity, and density, respectively.

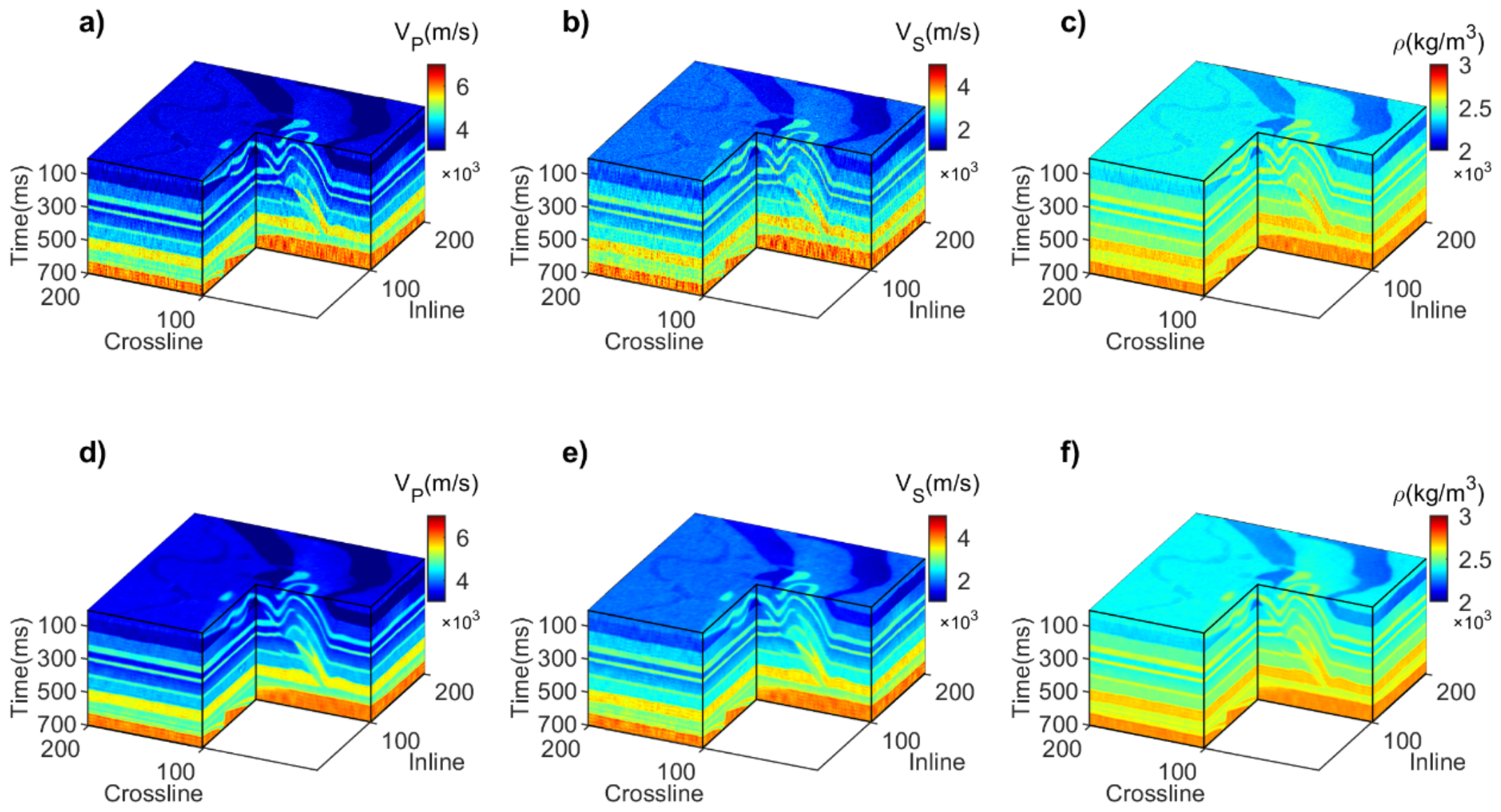

Figure 11. The estimated (a) P-wave velocity, (b) S-wave velocity, and (c) density by global algorithm without TV . The estimated (d) P-wave velocity, (e) S-wave velocity, and (f) density by global algorithm with L2-TV.

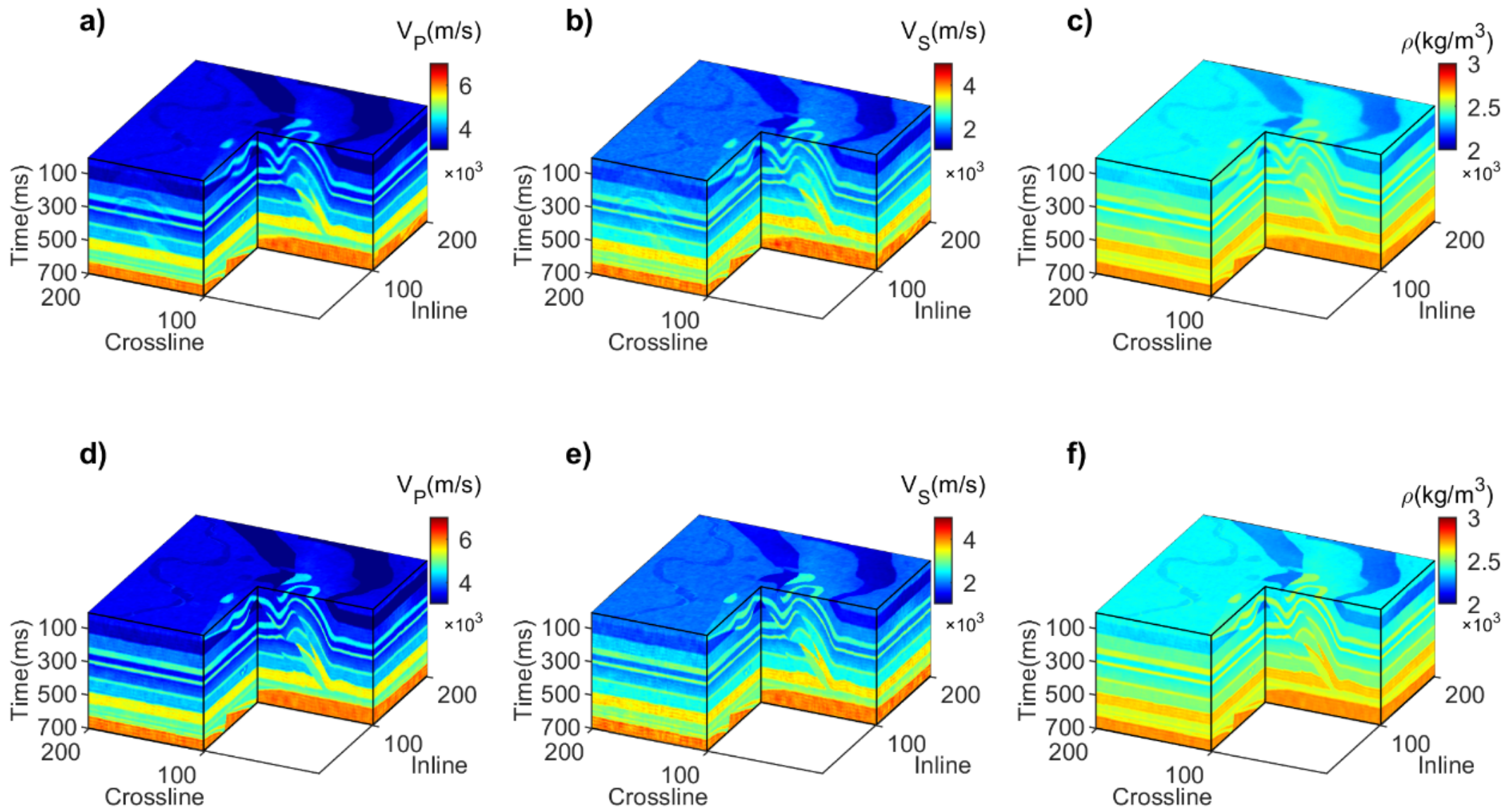

Figure 12. The estimated (a) P-wave velocity, (b) S-wave velocity, and (c) density by global algorithm with L1-TV . The estimated (d) P-wave velocity, (e) S-wave velocity, and (f) density by global algorithm with fault-guided L2.

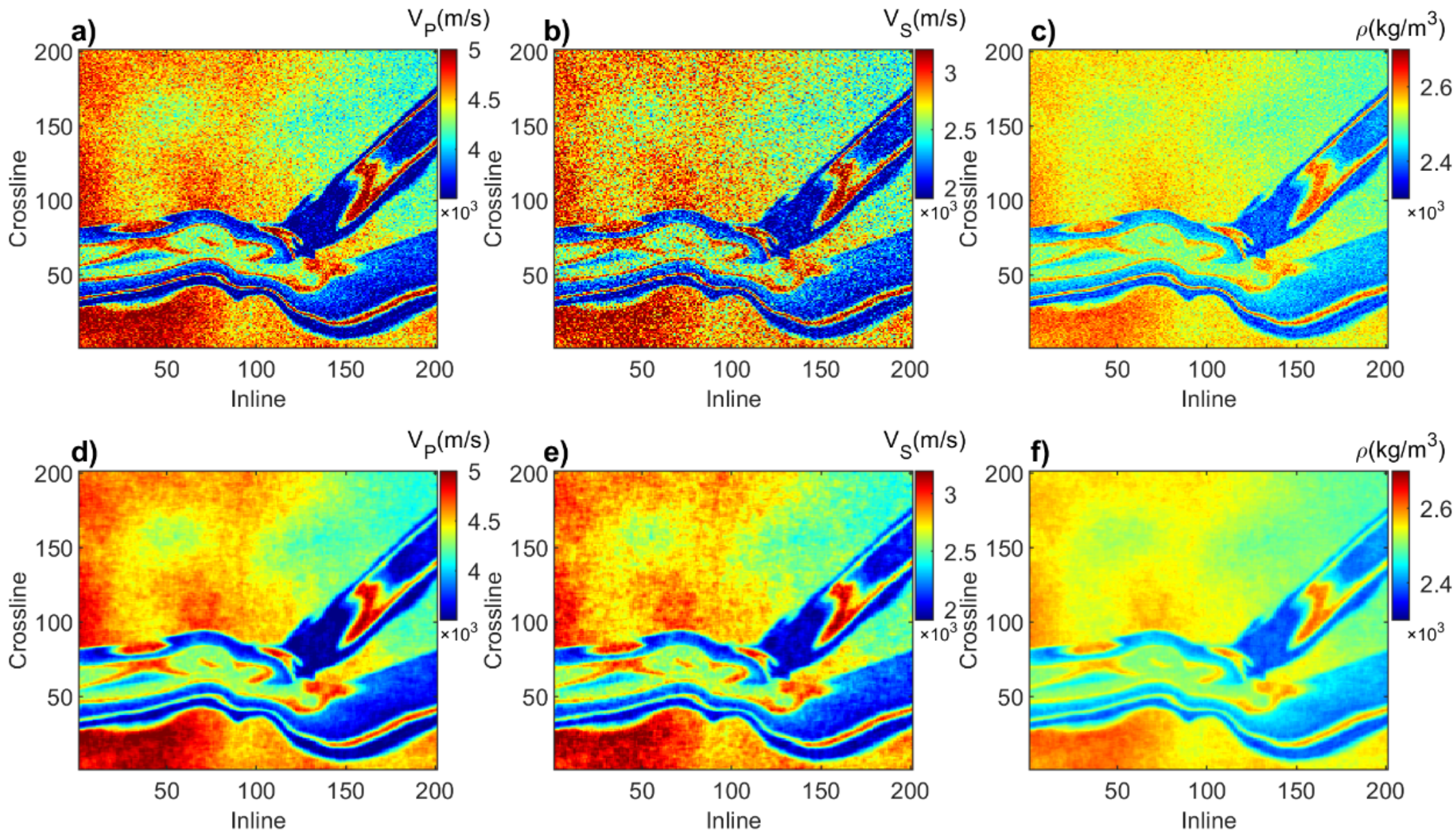

Figure 13. The 230 ms time slices of (a) P-wave velocity, (b) S-wave velocity, and (c) density by global algorithm without TV . The 230 ms time slices of (d) P-wave velocity, (e) S-wave velocity, and (f) density by global algorithm with L2-TV.

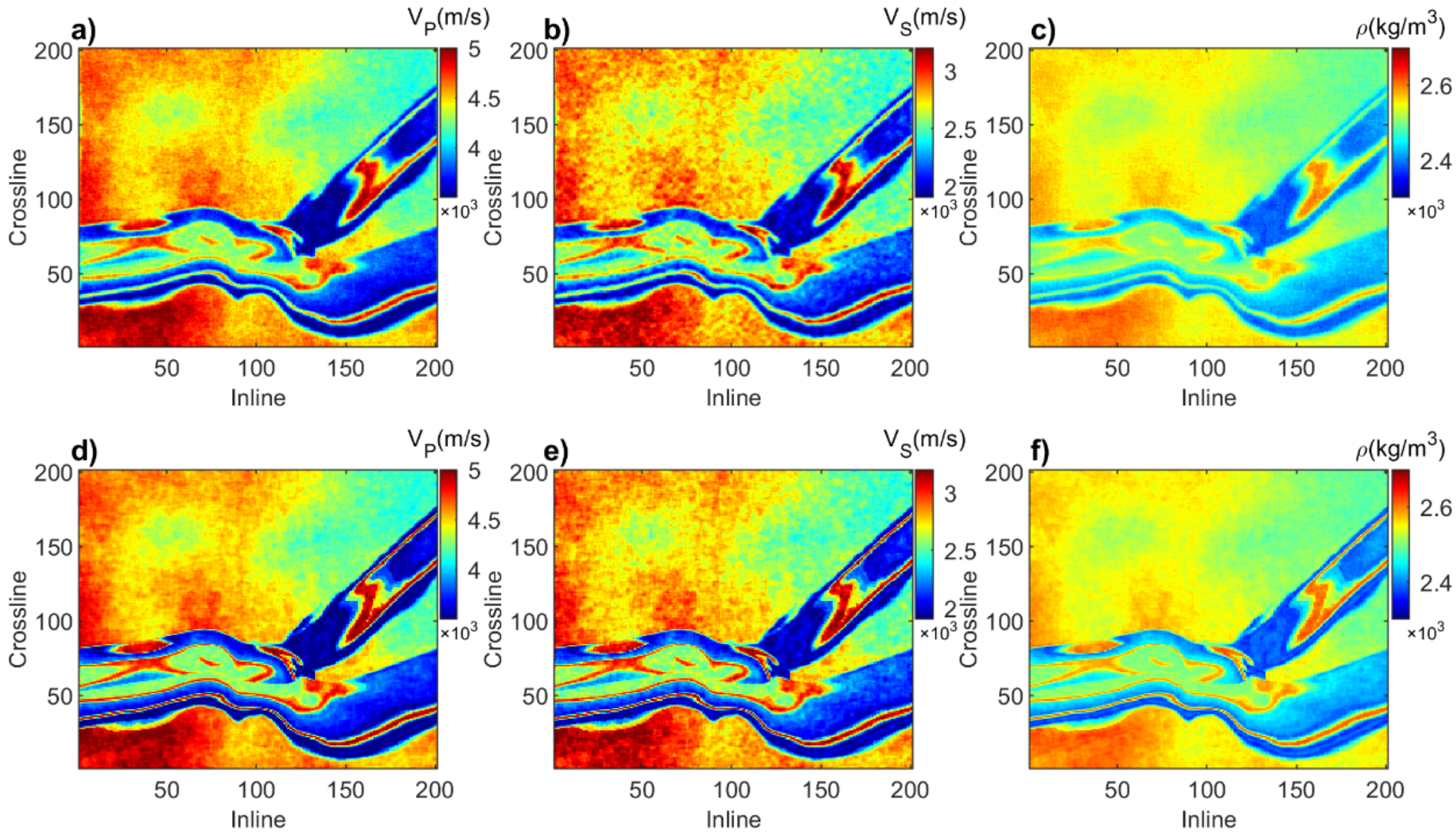

Figure 14. The 230 ms time slices of (a) P-wave velocity, (b) S-wave velocity, and (c) density global algorithm with L1-TV . The 230 ms time slices of (d) P-wave velocity, (e) S-wave velocity, and (f) density by global algorithm with fault-guided L2.

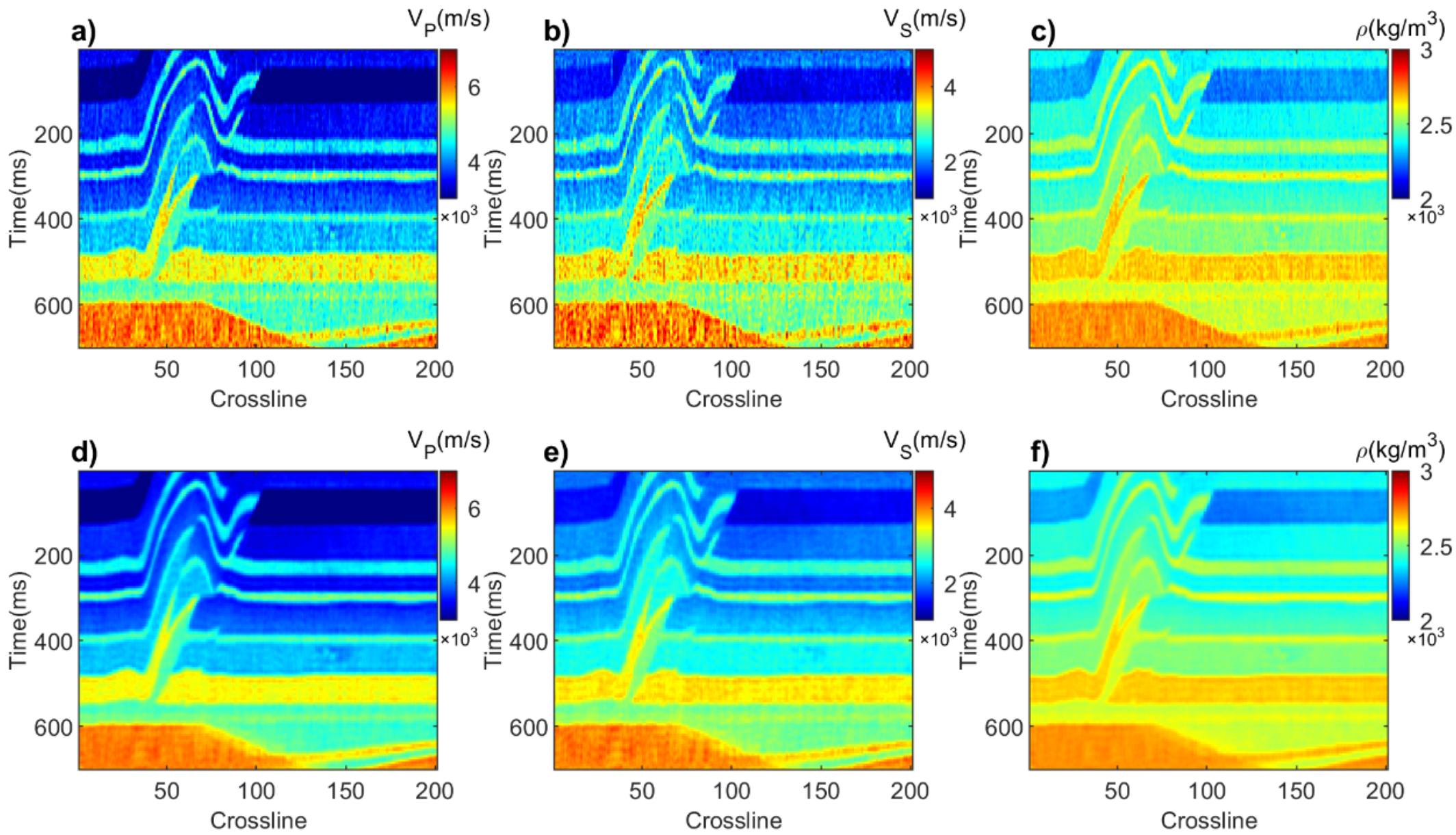

Figure 15. The estimated (a) P-wave velocity, (b) S-wave velocity, and (c) density by global algorithm without TV at Inline = 100; The estimated (d) P-wave velocity, (e) S-wave velocity, and (f) density by global algorithm with L2-TV at Inline = 100.

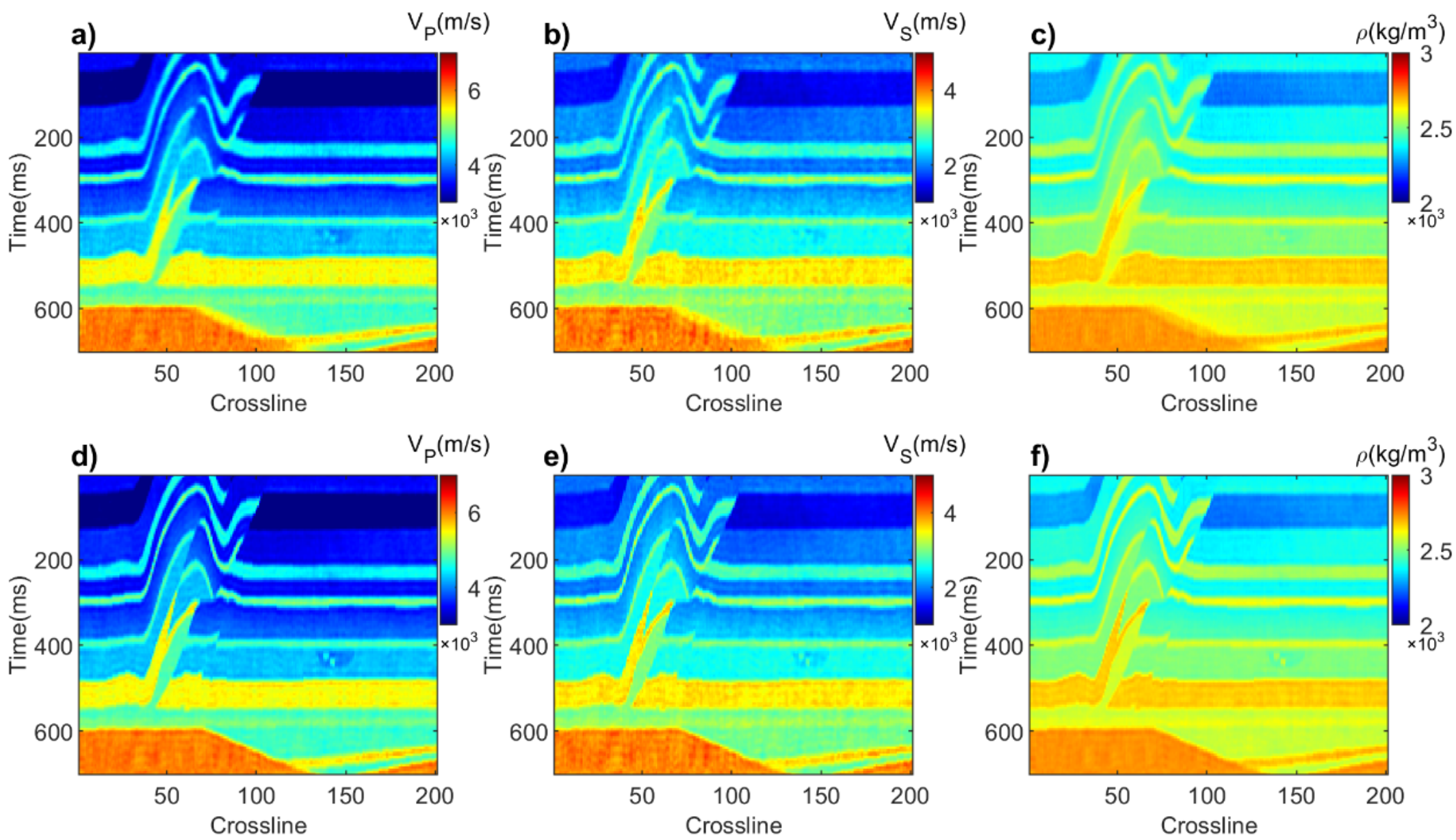

Figure 16. The estimated (a) P-wave velocity, (b) S-wave velocity, and (c) density by global algorithm with L1-TV at Inline = 100. The estimated (d) P-wave velocity, (e) S-wave velocity, and (f) density by global algorithm with fault-guided L2 at Inline = 100.

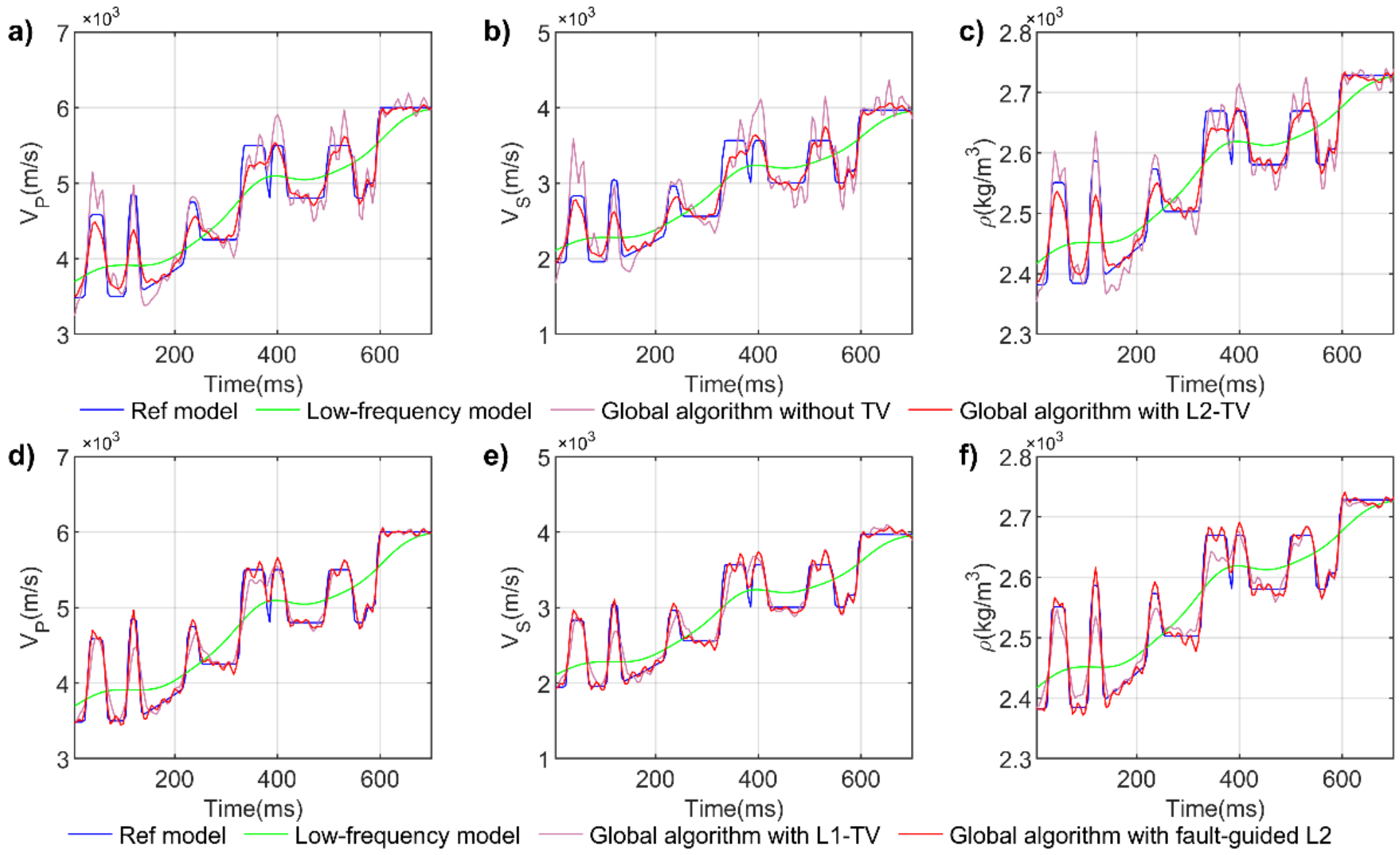

Figure 17. The estimated (a) P-wave velocity, (b) S-wave velocity, and (c) density by global algorithm without TV and global algorithm with L2-TV at Inline = 100 and Crossline = 50 . The estimated (d) P-wave velocity, (e) S-wave velocity, and (f) density by global algorithm with fault-guided L2 at Inline = 100 and Crossline = 50.

## FIELD DATA EXAMPLE

Another field dataset is used to evaluate the robustness of the GSCI. The dataset contains 965 crosslines and 1191 inlines. The seismic angle gathers at 9°, 19°, and 31° are presented in Figure 18. Figure 19 displays the 3D P-wave velocity, S-wave velocity, and density models inverted using the conventional algorithm and the GSCI, while Figure 20 provides the corresponding well-tie sections together with the well logs from Well 1, Well 2, and Well 3.

The inversion results at the well locations are further compared. Figure 21 presents the P-wave velocity inversion results for Well 3, Well 2, and Well 1 (Figures 21a–21c). In each panel, the blue curve represents the well log filtered to a 55 Hz seismic bandwidth, the green dashed line denotes the low-frequency model, the black curve corresponds to the inversion result from the conventional algorithm, and the red curve represents the result obtained using the GSCI algorithm. The P-wave velocity results from the two methods are generally consistent. The S-wave velocity inversion results are compared in Figure 22 using the same curve definitions, showing that the GSCI produces noticeably improved S-wave velocity estimates. The density inversion results are presented in Figure 23, where the density obtained from the GSCI shows significantly better agreement with the well log curves than that from the conventional method.

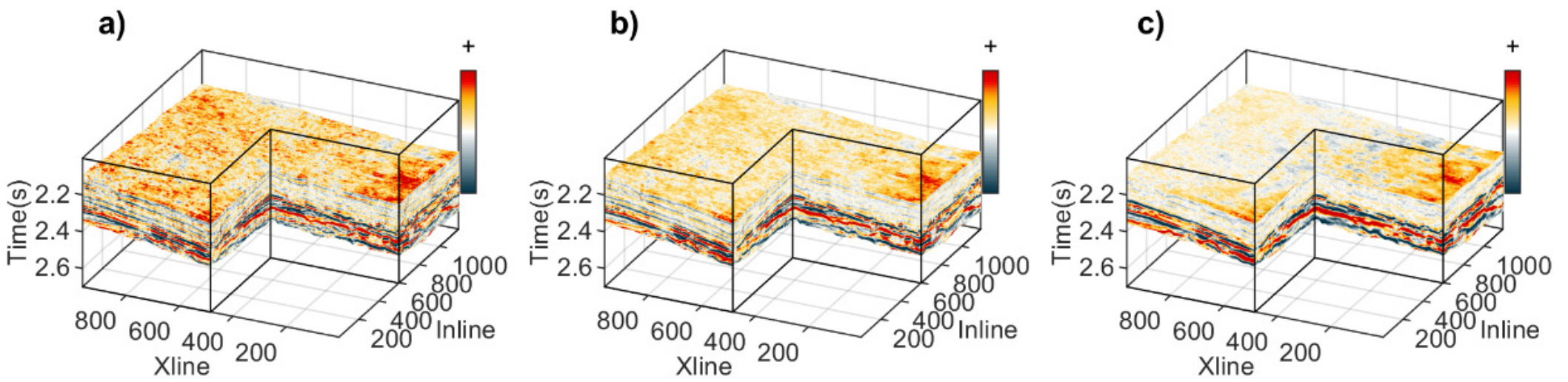

Figure 18. The 3D field prestack gathers with the angles of (a) 9°, (b) 19°, and (c) 31°. Wells 1, 2, and 3 are located in the study area, with a 2 ms sampling interval.

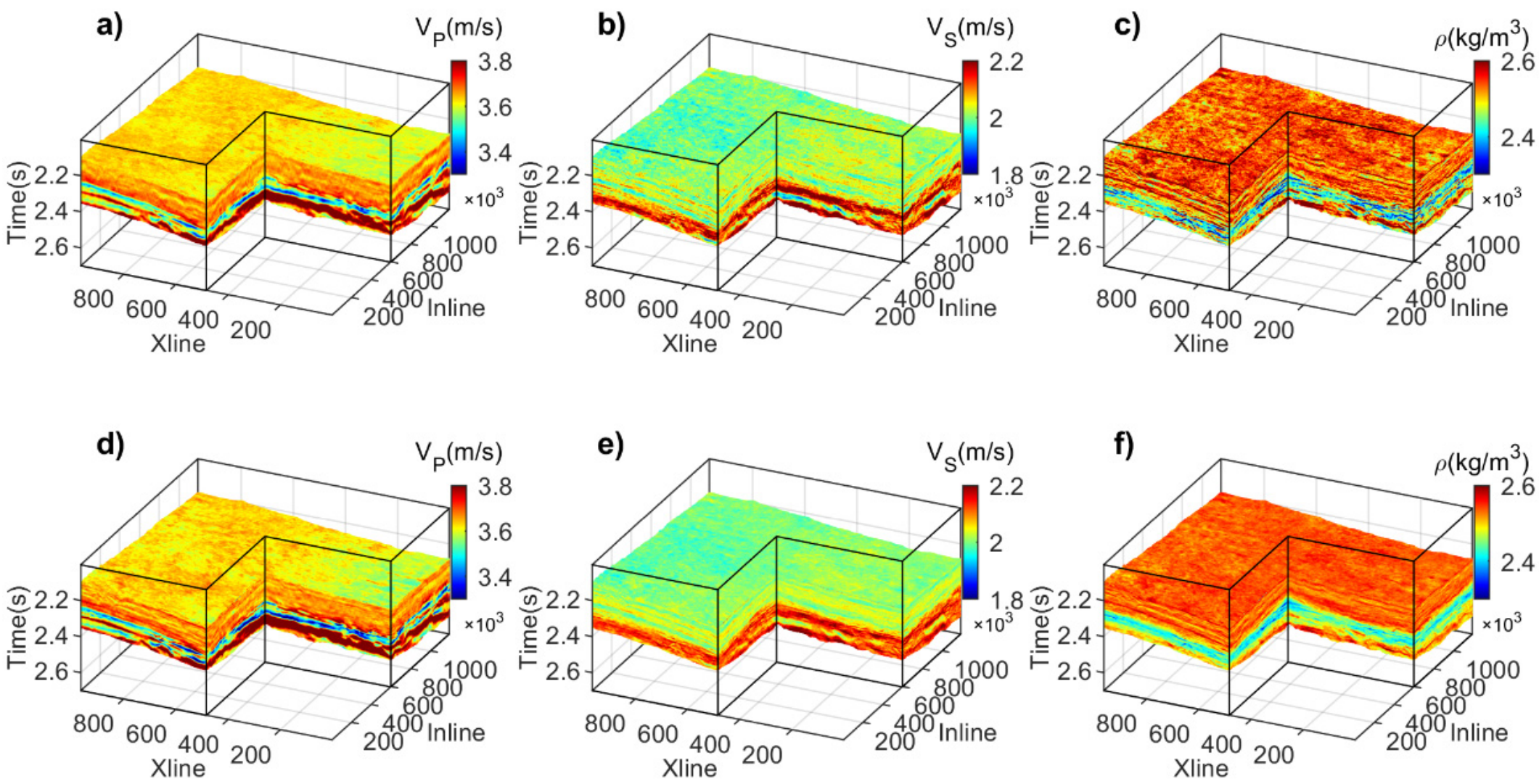

Figure 19. The estimated (a) P-wave velocity, (b) S-wave velocity, and (c) density by regular algorithm. The estimated (d) P-wave velocity, (e) S-wave velocity, and (f) density by GSCI.

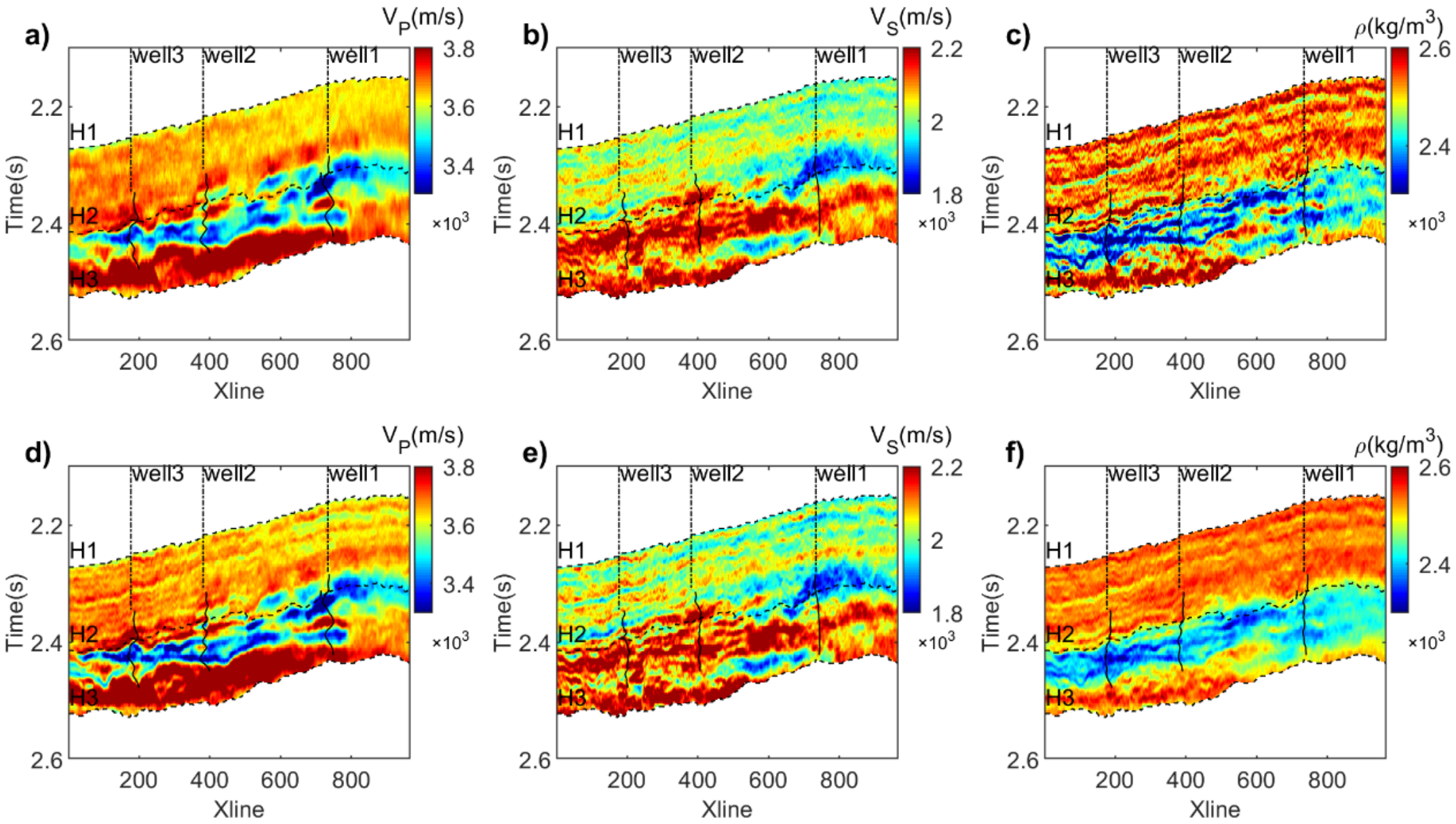

Figure 20. The estimated (a) P-wave velocity, (b) S-wave velocity, and (c) density interwell profiles by regular algorithm. The estimated (d) P-wave velocity, (e) S-wave velocity, and (f) density interwell profiles by GSCI.

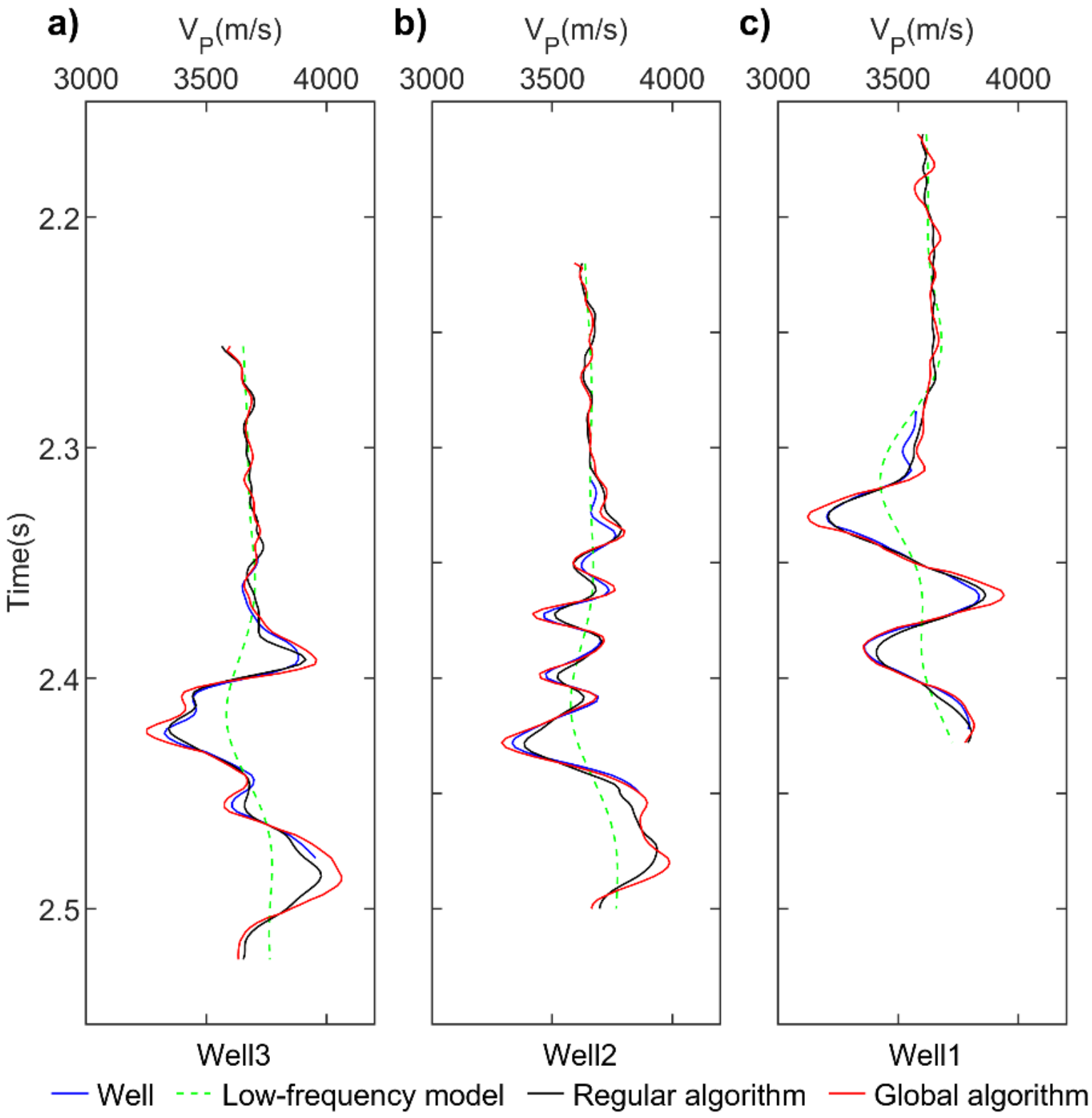

Figure 21. Estimated P-wave velocity logs at (a) Well 3, (b) Well 2, and (c) Well 1. Each panel compares the well log (blue), the low-frequency model (green dashed), and the P-wave velocity

inversion results from the regular algorithm (black) and the GSCI (red). The well log is obtained by smoothing the true log curves to the seismic bandwidth (55 Hz).

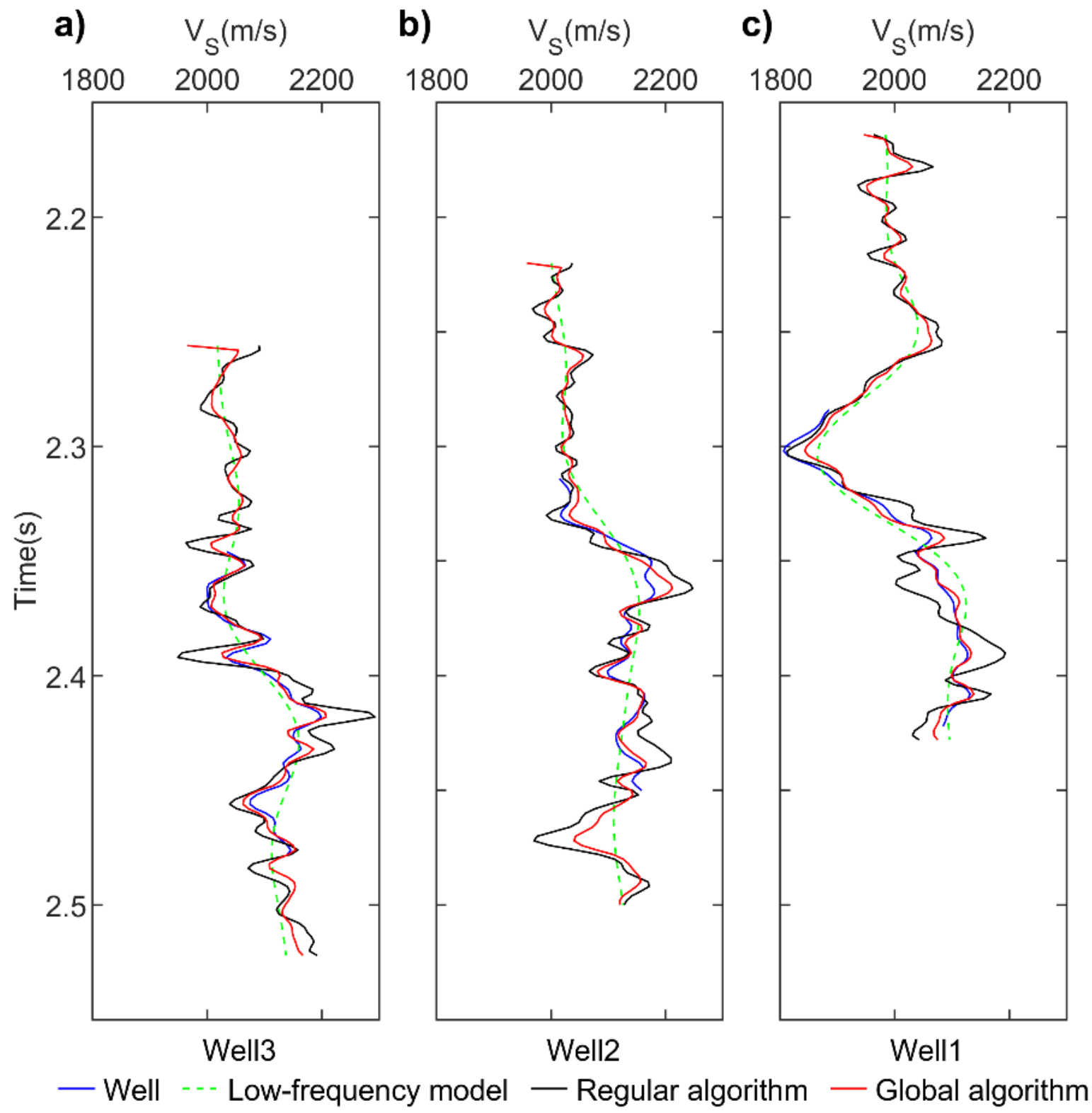

Figure 22. Estimated S-wave velocity logs at (a) Well 3, (b) Well 2, and (c) Well 1. Each panel compares the well log (blue), the low-frequency model (green dashed), and the S-wave velocity inversion results from the regular algorithm (black) and the GSCI (red). The well log is obtained by smoothing the true log curves to the seismic bandwidth (55 Hz).

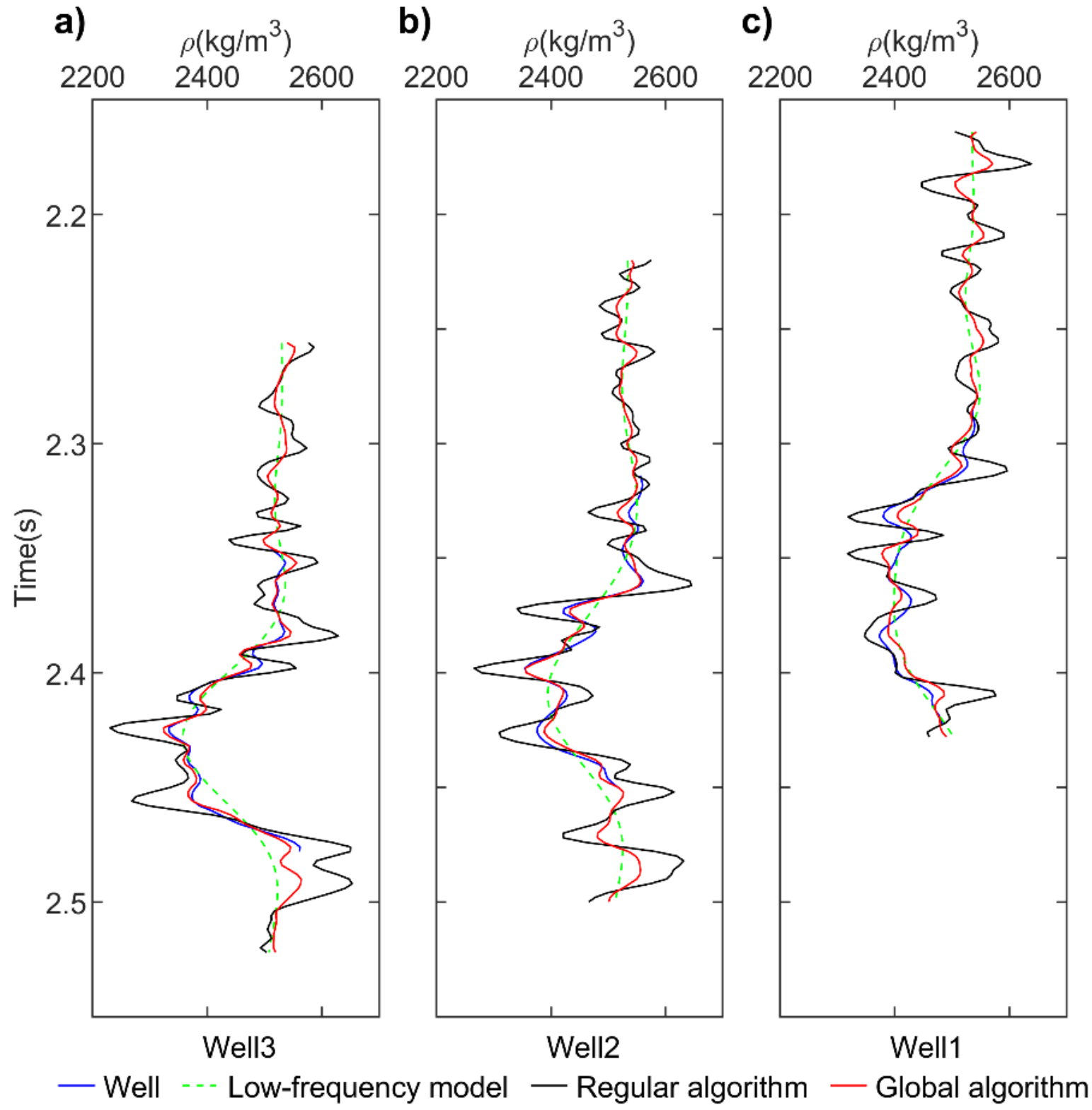

Figure 23. Estimated density logs at (a) Well 3, (b) Well 2, and (c) Well 1. Each panel compares the well log (blue), the low-frequency model (green dashed), and the density inversion results from the regular algorithm (black) and the GSCI (red). The well log is obtained by smoothing the true log curves to the seismic bandwidth (55 Hz).

## DISCUSSION

The development of the proposed inversion framework was initially motivated by the need to incorporate structural constraints more effectively into pre-stack AVO inversion. During the formulation of the method, it became apparent that seismic inversion is intrinsically a multidimensional problem in which different dimensions affect different aspects of the inversion results. Representing the elastic parameters as a four-dimensional tensor therefore provides a natural and unified formulation for describing the inversion model. In this framework, the spatial dimensions characterize the three-dimensional distribution of subsurface media, while the parameter dimension represents the coupling among elastic parameters such as P-wave velocity, S-wave velocity, and density. This formulation allows the inversion problem to be analyzed and constrained along different dimensions, making it possible to design targeted regularization strategies that improve stability, suppress noise amplification, and preserve spatial continuity.

From a theoretical perspective, the tensor-based formulation also differs fundamentally from conventional trace-by-trace or section-based inversion strategies. Conventional approaches often decompose a 3D inversion problem into a collection of independent lower-dimensional problems, which may neglect the spatial correlations among neighboring traces. In contrast, the multidimensional tensor representation treats the inversion as a global problem and preserves the intrinsic coupling between spatial locations and elastic parameters. As a result, structural continuity and lateral variations can be incorporated more naturally into the inversion framework, leading to more geologically consistent solutions.

Another important aspect of this study is the incorporation of geological structural information. Conventional structural constraints are commonly derived from structure tensors computed directly from seismic amplitudes, which provide local estimates of dip and azimuth. Although such approaches can enhance structural continuity, they primarily describe local geometric trends and may not fully capture the interpreted geological framework. In contrast, the GSCI integrates structural information obtained from seismic interpretation, including horizons and faults. Horizon constraints are introduced through layer-flattening techniques to improve stratigraphic continuity, while fault-guided regularization is used to preserve structural discontinuities across fault zones. This interpretation-driven strategy enables the inversion process to follow the actual geological architecture more closely.

The numerical experiments and field data application demonstrate that the combination of multidimensional tensor formulation and interpretation-based structural constraints improves the stability and geological consistency of the inversion results. The inverted elastic parameters exhibit clearer structural boundaries and better agreement with well log data compared with conventional approaches. Nevertheless, the effectiveness of the structural constraints depends on the reliability of seismic interpretation, and uncertainties in the interpreted horizons or faults may influence the inversion results. Future research may therefore focus on improving the robustness of the structural constraints and extending the framework to more complex geological scenarios, such as anisotropic media and fluid-sensitive inversion.

## CONCLUSION

This study proposes a global pre-stack AVO inversion method that combines multidimensional matrix multiplication, multidimensional regularization, and structural constraints to improve elastic parameter estimation in complex geological environments. The framework represents model parameters as a 4D tensor and constructs the forward modeling process using multidimensional matrix multiplication, providing an efficient formulation for global inversion of 3D seismic data. Multidimensional regularization is incorporated to enhance inversion stability and suppress noise effects, while geological constraints derived from interpreted horizons and faults improve structural consistency and boundary preservation.

Both synthetic experiments, including the Marmousi-2 and 3D overthrust models, and applications to field seismic data demonstrate that the GSCI achieves improved inversion accuracy, better lateral continuity, stronger noise robustness, and closer agreement with well log data compared with conventional approaches. These results indicate that the GSCI provides a robust and efficient approach for elastic parameter inversion in structurally complex regions and offers promising potential for reservoir characterization in hydrocarbon exploration.

## ACKNOWLEDGMENTS

## DATA AND MATERIALS AVAILABILITY

Data associated with this research are available and can be obtained by contacting the corresponding author.

## APPENDIX A

## MULTIDIMENSIONAL TENSOR OPERATIONS

Based on basic matrix theory, multidimensional tensor operation notations are introduced to simplify the derivation of the proposed inversion framework. Figure A-1 illustrates the multiplication of four different matrices along the four dimensions of a four-dimensional (4D) tensor.

The conventional matrix multiplication corresponds to the operation along the Z-dimension of the 4D tensor (Figure A-1a). Multiplication along the X-dimension is shown in Figure A-1b, while multiplication along the Y-dimension is illustrated in Figure A-1c. Finally, multiplication along the parameter dimension (i.e., the fourth dimension) of the 4D tensor is presented in Figure A-1d. The corresponding mathematical expressions for these four cases are provided in the preceding section.

Tensor multiplication can be denoted using the symbol × with a directional subscript to indicate the operating dimension. In practical applications, this notation can also be simplified according to the physical meaning of the operation.

**Definition 1:** Calculation rules for the multiplication of a 4D tensor and a 2D matrix.

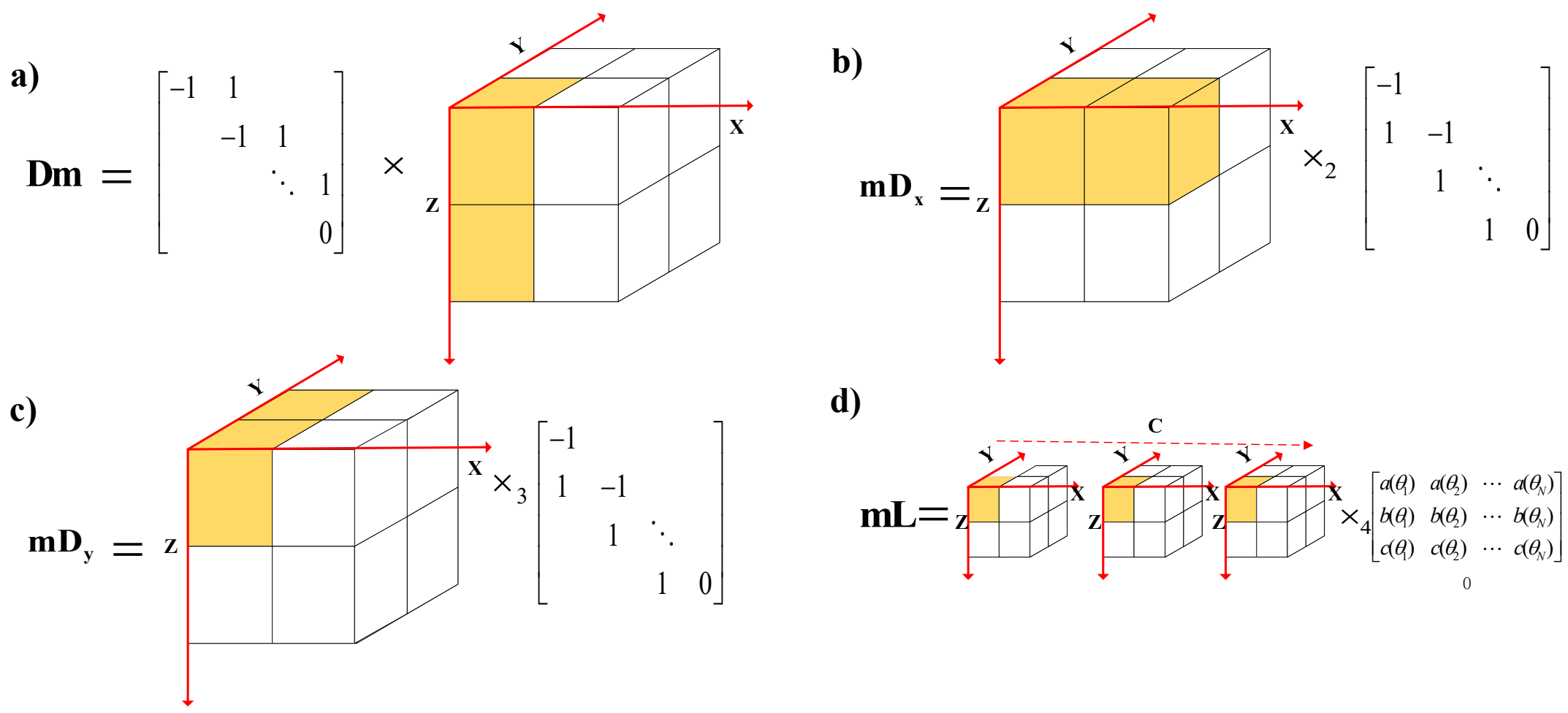

Figure A-1. Schematic diagram illustrating the calculation rules for the multiplication between a 4D tensor and a 2D matrix. (a) Differencing along the Z-dimension of the 4D tensor using a Z-direction difference matrix. (b) Differencing along the X-dimension of the 4D tensor using an X-direction difference matrix. (c) Differencing along the Y-dimension of the 4D tensor using a Y-direction difference matrix. (d) Calculation along the parameter dimension of the 4D tensor using a parameter-direction coefficient matrix.

# APPENDIX B

## DIMENSION-WISE INVERSION

Assume that the objective function is defined as

$$\hat{\mathbf{m}} = \arg\min_{\mathbf{m}} \| \mathbf{d} - \mathbf{G} \times \mathbf{m} \times_4 \mathbf{L} \|_2^2 + \alpha_1 \| \mathbf{m} \times_2 \mathbf{D_x} \|_2^2 \quad \text{(B-1)}$$

To obtain the optimal solution, the derivative of the objective function with respect to the model parameters is computed and set to zero, which leads to

$$\mathbf{G'G} \times \mathbf{m} \times_4 \mathbf{LL'} + \alpha_1 \mathbf{m} \times_2 \mathbf{D_x D_x'} = \mathbf{G'} \times \mathbf{d} \times_4 \mathbf{L'} \quad \text{(B-2)}$$

When $\alpha_1 << 1$, equation B-2 can be simplified as

$$\mathbf{G'G} \times \mathbf{m} \times_4 \mathbf{LL'} + \alpha_1 \mathbf{m} \times_2 \mathbf{D_x D_x'} \approx \mathbf{G'G} \times \mathbf{m} \times_4 \mathbf{LL'} + \mathbf{G'G} \times \mathbf{m} \times_4 \mathbf{LL'} \times_2 \alpha_1 \mathbf{D_x D_x'} \quad \text{(B-3)}$$

The solution of equation B-3 can therefore be written as

$$\hat{\mathbf{m}} \approx \left[ \left[ \left( \mathbf{G'G} \right)^{-1} \mathbf{G'} \right] \mathbf{d} \times_4 \left[ \mathbf{L'} \left( \mathbf{LL'} \right)^{-1} \right] \right] \times_2 \left[ \left( \mathbf{I} + \alpha_1 \mathbf{D_x D_x'} \right)^{-1} \right] \quad \text{(B-4)}$$

This formulation provides the solution for the inversion along the X-dimension. A similar derivation can be obtained for the Y-dimension following the same procedure. When $\alpha_1, \alpha_2 << 1$ and $p = 2$, equation (22) can be approximated by equation (23), which gives the dimension-wise solution used in the proposed inversion framework.